%% file: main.tex
\documentclass[prx,twocolumn,unsortedaddress,superscriptaddress]{revtex4-2}
\usepackage[english]{babel}
\addto\captionsenglish{}
\usepackage{graphicx}

\usepackage[caption=false,font=footnotesize]{subfig}

\usepackage{epsfig,color}
\usepackage{xcolor}
\usepackage{amsmath,amssymb,eufrak,calc}
\usepackage{amsthm}
\usepackage{enumerate}
\usepackage{hyperref}
\usepackage{amsxtra,amsfonts,bm,txfonts}
\usepackage{blkarray}
\usepackage{adjustbox}
\usepackage{braket}
\usepackage{adjustbox}
\usepackage{xcolor}
\usepackage{tikz-feynman}
\usepackage{tikz}
\usepackage{tikz-cd}

\newcommand{\fulleqref}[1]{Eq. \eqref{#1}} 
\newcommand{\paren}[1]{\left( {#1} \right)} 

\newcommand{\abs}[1]{\left|{#1}\right|}

\usepackage{siunitx}

\begin{document}

\title{Quantum Simulation of non-Abelian Lattice Gauge Theories:\\ a variational approach to $\mathbb{D}_8$}
\date{\today}

\input{Sections/authors_abstract}

\maketitle

\section{Introduction}
\input{Sections/Introduction}

\input{Sections/Background}


\section{The variational procedure}\label{variational}

In this section, we briefly summarize the basic ideas behind the algorithm we propose for quantum simulation of the non-Abelian gauge models under consideration, based on the variational principle. We also outline possible implementation strategies on a specific quantum hardware, trapped-ion qudit quantum computers like the one developed in Ref.~\cite{ringbauer2022universal}.

Variational algorithms have emerged as a promising candidate to implement subroutines for resource-efficient quantum simulation~\cite{Cerezo2021}. In this approach, a variational Ansatz circuit is chosen and efficiently prepared on the quantum device, and a hybrid quantum-classical feedback loop is established. On the quantum device, a set of observables is measured and given as an input to the classical co-processor. The latter employs a minimization procedure for a pre-defined cost function, and calculates the updated parameters of the parametrized Ansatz, which are given as input to the quantum device. This procedure is repeated until convergence to an optimal set of parameters is reached. The variational approach can provide an efficient parametrization of many-body states, for example ground and excited states of a Hamiltonian or even time-evolved states under some unitary evolution. Nevertheless, variational algorithms encounter several difficulties, some of which are due to the current state of quantum technology, while others are more fundamental in nature. Examples of such difficulties are  low robustness towards noise~\cite{PhysRevResearch.5.023025}, overhead in terms of establishing the quantum-classical feedback~\cite{Hoefler2023}, susceptibility to local minima and the encounter of barren plateaus~\cite{McClean2018}.

Nevertheless, the variational approach has been applied successfully in numerical simulations in the context of quantum simulation of lattice gauge theories in the past, mostly for systems with Abelian gauge symmetry~\cite{Ferguson2021,paulson_simulating_2021,zhang2023simulating,popov2024variational,Chan2024measurementVQE}, but also for some non-Abelian toy models~\cite{atas_su_2021,Atas2023chromo}. In the laboratory, variational protocols have been employed for simulating gauge theory models on a (1+1)-D~\cite{kokail_self_2019} and recently on a (2+1)-D lattice~\cite{meth2024simulating2d}. In the following, we elaborate on the specific variational algorithm used in this work, the variational quantum imaginary time evolution  (VarQITE) algorithm~\cite{Yuan2019}.

\subsection{The VarQITE algorithm}

One can formally write the equation of motion for a quantum state described by a density matrix $\rho(\tau)$ as a function of the (imaginary) time $\tau$ as 
\begin{equation}
    \frac{d}{d\tau}\rho(\tau) = \mathcal{L}[\rho(\tau)] \, .
\label{eq:EoM}
\end{equation}
For a Hamiltonian $H$, the super-operator reads
\begin{align}
    \mathcal{L}[\rho(\tau)]=-\{H, \rho(\tau)\}+2\langle H\rangle \rho(\tau), 
\label{eq:imag_time_superop}
\end{align}
where the average is $\langle\cdot\rangle\equiv \text{Tr}[\rho(\tau)\cdot]$ and the anti-commutator $\{A,B\} \equiv AB+BA$. The second term on the right-hand side of Eq.~\eqref{eq:imag_time_superop} takes care of the normalization of the density matrix, i.e., $\text{Tr}[\rho(\tau)] \equiv 1$.

In case we have a parametrized density matrix, $\rho(\tau) = \rho(\boldsymbol{\theta}(\tau))$ for a set of parameters $\boldsymbol{\theta}$, solving Eq.~\eqref{eq:EoM} can be cast as a minimization problem; solutions are obtained by minimizing the so-called McLachlan distance,
\begin{align}
    \mathcal{D(\boldsymbol{\theta})} \equiv \Big\Vert \sum_{\mu}\frac{\partial \rho}{\partial\theta_{\mu}}\dot{\theta}_{\mu}-
    \mathcal{L}[\rho]\Big\Vert^2 \,,
\end{align}
with respect to trajectories of the parameters $\boldsymbol{\theta}(\tau)$. Furthermore, the minimal action principle teaches us that one can formulate equivalent equations of motion for the parameters,
\begin{align}
    \sum_{\nu}M_{\mu\nu}\dot{\theta}_{\nu}(\tau) = V_{\mu}\,,
\label{eq:eom_for_the_thetas}
\end{align}
where the matrix $M_{\mu\nu}$ is the Fubini--Study metric tensor,
\begin{align}
M_{\mu\nu} = &\mathrm{Tr}\Bigg[\frac{\partial \rho}{\partial\theta_{\mu}}\frac{\partial \rho}{\partial\theta_{\nu}}\Bigg] \,
\label{eq:metric_tensor}
\end{align}
and the vector $V_{\mu}$ is defined as
\begin{align}
V_{\mu} = \mathrm{Tr}\Bigg[\frac{\partial \rho}{\partial\theta_{\mu}}\mathcal{L}[\rho]\Bigg].
\label{eq:vector_V}
\end{align}

In this work, we consider a family of trial states given by a parametrized quantum circuit (PQC), as follows:
\begin{align}
    |\psi(\boldsymbol{\theta})\rangle=\mathcal{U}(\boldsymbol{\theta})\ket{\psi_0}=U_{N}\left(\theta_{N}\right) \ldots U_{k}\left(\theta_{k}\right) \ldots U_{1}\left(\theta_{1}\right)\ket{\psi_0}\,,
    \label{eq:TrialState}
\end{align}
where $\ket{\psi_0}$ is an initial state and $U_k(\theta_k)$ are unitaries that, acting on the initial state, generate the PQC. With the identification $\rho(\boldsymbol{\theta}) \equiv \ket{\psi(\boldsymbol{\theta})}\bra{\psi(\boldsymbol{\theta})}$, one obtains the matrix $M_{\mu\nu}$ and the vector $V_{\mu}$ as follows:
\begin{align}
    M_{\mu\nu} = \text{Re}(\langle\partial_{\mu}\psi|\partial_{\nu}\psi\rangle) - \langle\partial_{\mu}\psi|\psi\rangle\langle\psi|\partial_{\nu}\psi\rangle
\label{eq:Mmunu}
\end{align}
and
\begin{align}
    V_{\mu} = \langle\partial_{\mu}\psi| H |\psi\rangle + \langle\psi| H |\partial_{\mu}\psi\rangle = \partial_{\mu}\langle H \rangle. 
\label{eq:vector_V_imag}
\end{align}
As long as these quantities depend on the point in the parameter space $\boldsymbol{\theta}$, Eq.~\eqref{eq:eom_for_the_thetas} is non-linear and therefore cannot be solved efficiently without resorting to the quantum device. In Ref.~\cite{popov2024variational} one can find a protocol for measuring these quantities on a qudit device. Here, we focus on the formulation of the PQC for the ground-state preparation of models with $\mathbb{D}_8$ discrete non-Abelian symmetry.

\subsection{Parametrized quantum circuit for qudits}
While the VarQITE algorithm, due to its versatility, holds the promise for implementation on various quantum devices based on qubits as well on qudits, in this work we focus on a specific platform---trapped-ion qudit quantum hardware. This makes it possible to define qudit-tailored variational circuits, composed of the gates that are native to the ions. These are single qubit and qudit (acting on two of the $D$ levels) operations such as X and Y rotations
\begin{align}
    U^{i,j}_{X/Y}(\theta) = \exp\big(-i\theta\sigma^{i,j}_{X/Y}\big)
\end{align}
with 
\begin{align}
    (\sigma^{i,j}_X)_{m,n} &= \begin{cases}
    1\,, \quad \text{if }  (m,n) = (i,j) \text{ or } (j,i)  \\
    0\,, \quad \text{otherwise}
    \end{cases}\\
    (\sigma^{i,j}_Y)_{m,n} &= \begin{cases}
    -i \,, \quad \text{if }  (m,n) = (i,j)  \\
    i \,, \quad \text{if }  (m,n) = (j,i)  \\    
    0\,, \quad \text{otherwise}
    \end{cases}\\
\notag
\label{eq:two_level_pauli_matrices}
\end{align}
and entangling qubit and qudit Mølmer–-Sørensen (MS) gates
\begin{align}
    MS^{i,j}(\theta) = \exp\left(-\frac{i\theta}{4} \left(\sigma_X^{i,j}\otimes\bold{1} + \bold{1}\otimes \sigma_X^{i,j}\right)^2 \right).
\end{align}
In Section~\ref{D8_Hilbert}, we highlighted the fact that our LGT-to-hardware encoding requires a mixed qubit--qudit quantum hardware. This can easily be achieved in the trapped-ion platform because of the possibility to choose two levels in the higher dimensional qudit for information processing. The corresponding two-level rotations then become the Pauli X and Y rotations and the MS gate acts on the two levels of the qubit. 

\begin{figure*}
    \centering
    \includegraphics[scale = 0.35]{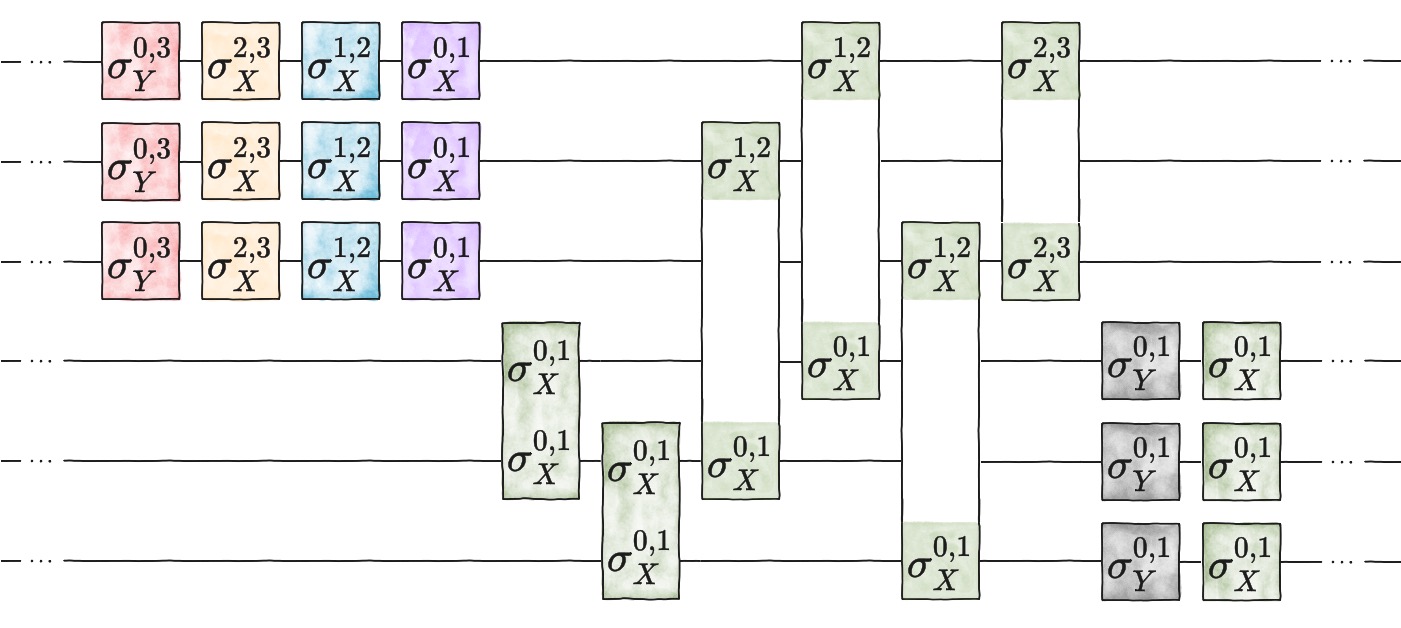}
    \caption{\textbf{Individual layer of the parametrized quantum circuit for the 1D system:} Each layer of the parametrized quantum circuit used to find the approximate ground state of the 1D system with non-Abelian $\mathbb{D}_8$-symmetry consists of two-level single-qudit rotations on ququarts, entangling two-body MS gates between neighbouring qubits, non-local MS gates between a qubit and a ququart, and single-qubit Pauli-$X$ and Pauli-$Y$ rotations. Each gate in the variational circuit is parametrized by a variational parameter $\theta$, chosen to be different in each layer. The software used for drawing the quantum circuits is taken from~\cite{drawio_library}. } 
    \label{fig:circuit_1D}
\end{figure*}

Specifically, we choose the variational circuit to be of a layered structure, with each layer containing the same type of gates but with a different variational parameter. The specific layers for the circuits used in this work are shown in FIG.~\ref{fig:circuit_1D} and FIG.~\ref{fig:circuit_pseudo_2D}.
It is worth noting that the single-qudit two-level rotations, combined with the MS entangling gate, constitute a universal set of qudit gates. A subset of these operations, while not necessarily universal, can in practice produce a highly expressive variational ansatz, able to approximate the ground state of the non-Abelian gauge theory models under consideration, as we show below.

\section{The one dimensional case}
\label{sec:1d}
\input{figures/1d_figure}

\begin{figure*}[t!]
    \centering
\subfloat{\label{fig:1d_energy}\includegraphics[width=0.5\textwidth]{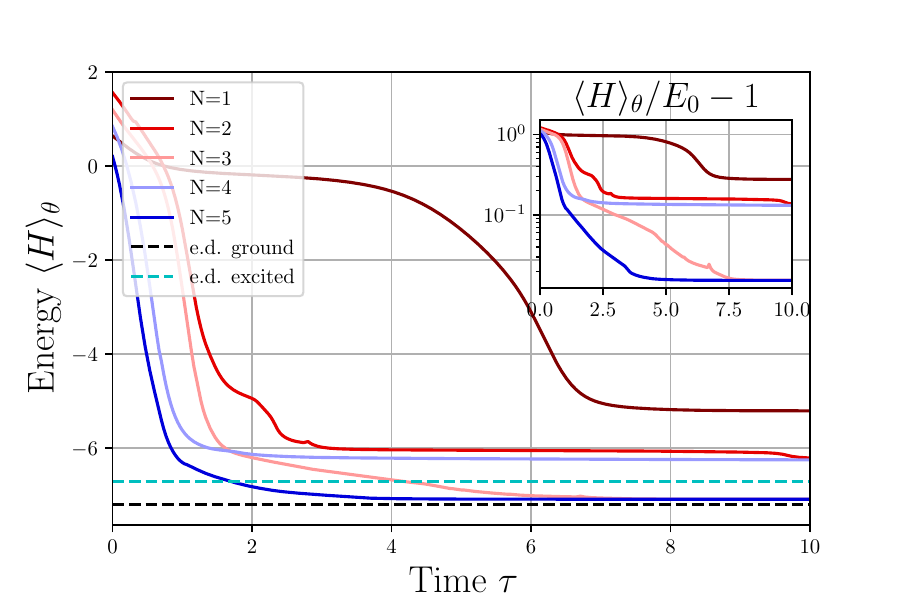}}%
\subfloat{\label{fig:1d_fidelity}\includegraphics[width=0.5\textwidth]{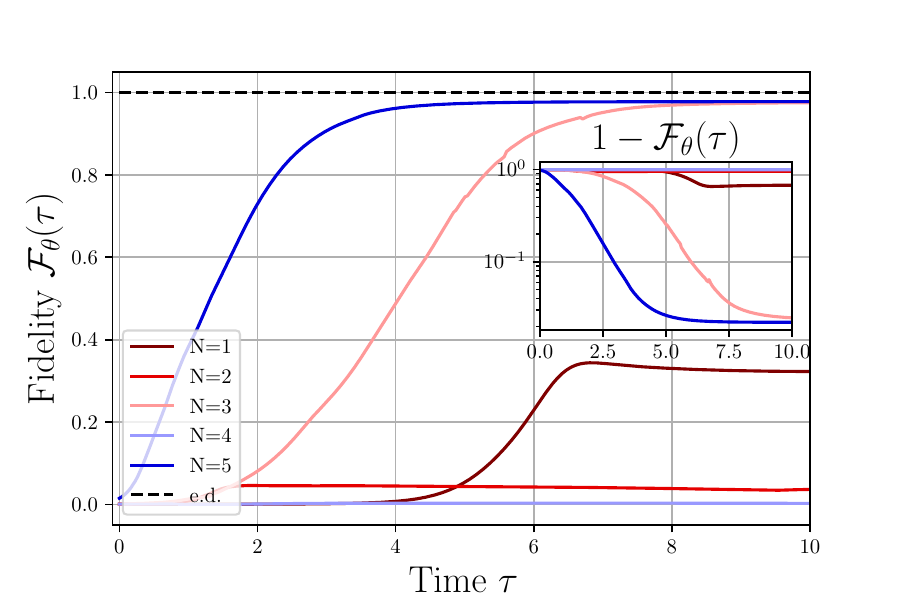}}%
    \caption{\textbf{VarQITE for the 1D system:} (a) The energy of the variational state $\ket{\psi(\boldsymbol{\theta)}}$ as a function of the imaginary time for $N = 1,2,3,4,5$ number of layers. The dashed black line shows the numerical value of the true ground state, whereas the cyan line the one of the numerical first excited state. Note that the latter is not gauge invariant and depends on the coefficient $\lambda$ of a penalty term introduced to the Hamiltonian, see Section \ref{sec:simulations_1D} (here, $\lambda = 0.1$). After initial fast convergence to low energies, the secondary slow optimization is primarily due to the penalty for Gauss' law violation. From the inset, one can deduce that after $\tau = 5$, the optimization practically stops, yielding a relative error $\sim1\%$ for $N = 5$ layers. (b) The fidelity of the variational state w.r.t.\ the true ground state from exact diagonalization as a function of the imaginary time $\tau$. As for the energy, the optimization practically stops after $\tau = 5$, leading to a fidelity of $\sim99\%$ for $N = 5$ layers.}
    \label{fig:results_12d}
\end{figure*}

\input{Sections/1d_analysis}

\section{The two dimensional case}

\input{Sections/2D_Case}

\section{Discussion and Summary} \label{discussion_and_summary}
\input{Sections/discussion_and_conclusions}

\section*{Acknowledgments}
E.Z. acknowledges  the support of the Israel Science Foundation (grant No. 374/24). 

ICFO-QOT group acknowledges support from:
European Research Council AdG NOQIA; MCIN/AEI (PGC2018-0910.13039/501100011033,  CEX2019-000910-S/10.13039/501100011033, Plan National FIDEUA PID2019-106901GB-I00, Plan National STAMEENA PID2022-139099NB, I00, project funded by MCIN/AEI/10.13039/501100011033 and by the “European Union NextGenerationEU/PRTR" (PRTR-C17.I1), FPI); QUANTERA MAQS PCI2019-111828-2;  QUANTERA DYNAMITE PCI2022-132919, QuantERA II Programme co-funded by European Union’s Horizon 2020 program under Grant Agreement No 101017733; Ministry for Digital Transformation and of Civil Service of the Spanish Government through the QUANTUM ENIA project call - Quantum Spain project, and by the European Union through the Recovery, Transformation and Resilience Plan - NextGenerationEU within the framework of the Digital Spain 2026 Agenda; Fundació Cellex; Fundació Mir-Puig; Generalitat de Catalunya (European Social Fund FEDER and CERCA program, AGAUR Grant No. 2021 SGR 01452, QuantumCAT \ U16-011424, co-funded by ERDF Operational Program of Catalonia 2014-2020); Barcelona Supercomputing Center MareNostrum (FI-2023-3-0024); Funded by the European Union (HORIZON-CL4-2022-QUANTUM-02-SGA  PASQuanS2.1, 101113690, EU Horizon 2020 FET-OPEN OPTOlogic, Grant No 899794, QU-ATTO, 101168628), ICFO Internal “QuantumGaudi” project; European Union’s Horizon 2020 program under the Marie Sklodowska-Curie grant agreement No 847648; “La Caixa” Junior Leaders fellowships, La Caixa” Foundation (ID 100010434): CF/BQ/PR23/11980043.

This project has received funding from the European Union’s Horizon Europe research and innovation programme under grant agreement No 101080086 NeQST. P.H. has further received funding from the QuantERA II Programme through the European Union’s Horizon 2020 research and innovation programme under Grant Agreement No 101017733, from the European Union under NextGenerationEU, PRIN 2022 Prot. n. 2022ATM8FY (CUP: E53D23002240006), from the European Union under NextGenerationEU via the ICSC - Centro Nazionale di Ricerca in HPC, Big Data and Quantum Computing. Views and opinions expressed are however those of the author(s) only and do not necessarily reflect those of the European Union or the European Commission. Neither the European Union nor the granting authority can be held responsible for them. This work was supported by
Q@TN, the joint lab between the University of Trento, FBK—Fondazione Bruno Kessler, INFN—National Institute for Nuclear Physics, and CNR—National Research Council.

P.P.P. acknowledges also support from the “Secretaria d’Universitats i Recerca del Departament de Recerca i Universitats de la Generalitat de Catalunya” under grant 2024 FI-3 00390, as well as the European Social Fund Plus.

E.G., P.P.P. and G.P. contributed equally to this work.
\section*{Appendix A: Matter removal transformation } \label{app_a}
\input{Sections/appendixA}



\bibliography{ref}

\end{document}

%% file: Sections/authors_abstract.tex
\author{Emanuele Gaz}
\address{Racah Institute of Physics, The Hebrew University of Jerusalem, Jerusalem 91904, Givat Ram, Israel.}

\author{Pavel P. Popov}
\address{ICFO - Institut de Ciencies Fotoniques, The Barcelona Institute of Science and Technology, Av. Carl Friedrich Gauss 3, 08860 Castelldefels (Barcelona), Spain}

\author{Guy Pardo}
\address{Racah Institute of Physics, The Hebrew University of Jerusalem, Jerusalem 91904, Givat Ram, Israel.}

\author{Maciej Lewenstein}
\address{ICFO - Institut de Ciencies Fotoniques, The Barcelona Institute of Science and Technology, Av. Carl Friedrich Gauss 3, 08860 Castelldefels (Barcelona), Spain}
\address{ICREA, Pg. Lluis Companys 23, 08010 Barcelona, Spain}

\author{Philipp Hauke}
\address{Pitaevskii BEC Center and Department of Physics, University  of  Trento,  Via Sommarive 14, I-38123 Trento, Italy}
\address{INFN-TIFPA, Trento Institute for Fundamental Physics and Applications, Trento, Italy}

\author{Erez Zohar}
\address{Racah Institute of Physics, The Hebrew University of Jerusalem, Jerusalem 91904, Givat Ram, Israel.}

\begin{abstract}
Quantum simulation of lattice gauge theories provides a powerful framework for understanding nonperturbative phenomena. 
In this work, we address the problem of a resource-efficient formulation of non-Abelian LGTs by focusing on the difficulty of simulating fermionic degrees of freedom  and the Hilbert space redundancy. First, we show a procedure that removes the matter and improves the efficiency of the hardware resources. We demonstrate it for the simplest non-Abelian group addressable with this procedure, $\mathbb{D}_8$, 
both in the cases of one (1D) and two (2D) spatial dimensions.
Then, with the objective of running a variational quantum simulation on real quantum hardware, we map the $\mathbb{D}_8$ lattice gauge theory onto qudit systems with local interactions. We propose a variational scheme for the qudit system with a local Hamiltonian, which can be implemented on a universal qudit quantum device as the one developed in \href{https://doi.org/10.1038/s41567-022-01658-0}{[Nat. Phys. 18, 1053 (2022)]}. 
Our results show the effectiveness of the matter-removing procedure, solving the redundancy problem and reducing the amount of quantum resources. 
This can serve as a way of simulating lattice gauge theories  in high spatial dimensions, with non-Abelian gauge groups, and including dynamical fermions.
\end{abstract}

%% file: Sections/Introduction.tex
Gauge theories are fundamental in describing the interactions among the elementary particles within quantum field-theory frameworks, particularly in high-energy physics \cite{quigg1997quantum}. 
Lattice Gauge Theories (LGTs) have become a pivotal tool for understanding the nonperturbative aspects.


These techniques provide a discrete spacetime lattice formulation that allows for the study of fundamental interactions \cite{bjorken_asymptotic_1969,gross_ultraviolet_1973}, particularly in regimes where perturbative methods fail, such as the low-energy dynamics responsible for quark confinement \cite{wilson_confinement_1974}. Traditional approaches to solving LGTs rely on Monte Carlo simulations, which have been widely successful in exploring properties of LGTs \cite{wilson_confinement_1974,kogut_hamiltonian_1975,kogut_introduction_1979}. However, these methods face significant limitations in addressing real-time dynamics and systems with finite chemical potential, due to the Euclidean time and the sign problem \cite{troyer_computational_2005}. As such, there has been a growing interest in leveraging new techniques to overcome these challenges. 

With the advent of quantum technologies, quantum simulation \cite{feynman_simulating_1982} has emerged as a promising alternative for exploring LGTs beyond the capabilities of classical methods \cite{wiese_ultracold_2013, zohar_quantum_2016}. 
By mapping the gauge and matter fields onto controllable quantum devices, it becomes possible to simulate real-time dynamics and tackle problems that are intractable using classical algorithms \cite{wiese_towards_2014}. Quantum simulation holds the potential to revolutionize the study of both static and dynamic properties of strongly-coupled gauge theories, especially in regimes where classical methods encounter severe limitations.

Various quantum simulation schemes have been proposed (see, e.g., the reviews \cite{wiese_towards_2014,zohar_quantum_2016,dalmonte_lattice_2016,banuls_review_2020,banuls_simulating_2020,klco_standard_2021,zohar_quantum_2022,aidelsburger_cold_2022,PRXQuantum.5.037001,halimeh2023coldatom}) and demonstrated across different platforms, from cold atoms \cite{schweizer_floquet_2019, mil_scalable_2020,yang_observation_2020,zhou_thermalization_2021} and trapped ions \cite{martinez_real-time_2016,kokail_self_2019,meth2024simulating2d} to superconducting qubits \cite{mezzacapo_non-abelian_2015, klco_quantum_2018, atas_su_2021,mildenberger_probing_2022}. Despite these advances, simulating LGTs in higher dimensions remains an outstanding challenge \cite{zohar_quantum_2022}, especially in the non-Abelian case. In $d>1$, the complexity of gauge theories (including their local symmetries), the many-body interactions \cite{kogut_hamiltonian_1975}, and the need to handle both fermionic and bosonic fields, imposes stringent requirements on simulation protocols. These factors contribute to limiting the existing experimental demonstrations of LGTs to specific scenarios.

In this paper, we reformulate the non-Abelian $\mathbb{D}_8$ LGT integrating out fermionic fields \cite{zohar_eliminating_2018, zohar_removing_2019}. 
The result is a simplified Hamiltonian that implements the fermionic statistics of matter through local interactions of bosonic degrees of freedom (qubits and qudits) and which removes the exponential redundancy in the simulated Hilbert space, thereby reducing the resource overhead required for simulation on quantum hardware. 
A previous analysis of an Abelian case was performed in Refs.~\cite{pardo2023resource, irmejs2023quantum}. This formulation maintains locality in the Hamiltonian even in higher spatial dimensions and avoids complications such as the Jordan--Wigner transformation \cite{jordan_uber_1928}. Previously, the quantum simulation of this group was studied theoretically by Refs.~\cite{alam2022,hardware_gonzalez,ballini2024symmetry}. The integrating-out of the fermions introduced here permits a significant economy of the employed degrees of freedom.

Following Ref.~\cite{popov2024variational}, we propose a variational quantum simulation protocol on a qudit trapped-ions platform \cite{ringbauer2022universal}. Unlike qubits, qudits use higher-dimensional Hilbert spaces, allowing for more efficient representations of gauge fields and matter, especially in the presence of non-Abelian gauge symmetries. 
This property facilitates the design of a resource-efficient variational simulation protocol on qudit quantum hardware, incorporating the complex structure of non-Abelian gauge theories.
By leveraging a hybrid quantum-classical algorithm, we can adaptively optimize quantum circuits and simulate both ground-state properties and dynamical evolution, circumventing the need for a precise engineering of the gauge-invariant Hamiltonian or an accurate Trotterization.

We test the performance of the variational algorithm in a numerical simulation in terms of ground-state preparation of a simple system with a non-Abelian gauge symmetry and benchmark it with respect to exact diagonalization results. We demonstrate the feasibility of our method, highlighting its potential for application to a broad range of problems in high-energy physics and beyond. Our study contributes to the ongoing effort to formulate a model of a non-Abelian gauge theory, simulable in the laboratory with existing technology, paving the way for future investigations into the rich physics of gauge theories in previously inaccessible regimes.

This paper is organized as follows. In Section \ref{LGT_review}, we briefly introduce the mathematical description of the target LGT focusing on $\mathbb{D}_8$ as a simple  non-Abelian case. This theory already shows the two features our formulation takes care of, the fermionic statistics of the matter and the redundancy of the Hilbert space. Then, in Sections \ref{method1} and \ref{method2} we introduce two unitary steps, $\mathcal{V}^{(1)}$ and $\mathcal{V}^{(2)}$, that allow us to eliminate the matter for $\mathbb{D}_8$ in arbitrary dimensions.
In Section \ref{variational}, we give an outline of the variational algorithm we use. Here, we present specific quantum circuits and discuss how the simulation protocol would be implemented on the quantum device. Next, we present the theoretical analysis of a pseudo-2D system (Section \ref{2d_pseudo_theory}) for which we use an independent exact analytical method for verification of the matter removal transformation. We later do the same for a single plaquette in Section \ref{2d_plaquette_theory}. We proceed to presenting a few numerical demonstrations (Section \ref{2d_simulations}) of the variational approach for these two systems. Finally, in section \ref{discussion_and_summary} we present our conclusions.

%% file: Sections/Background.tex
\section{The $\mathbb{D}_8$ lattice gauge theory} \label{LGT_review}
While lattice gauge theories were first introduced by Wilson in an Euclidean formalism \cite{wilson_confinement_1974}, the Hamiltonian formulation of LGTs has emerged as an alternative approach with several distinctive features \cite{kogut_hamiltonian_1975}. 
In the Hamiltonian approach, space is discretized on a $d$ dimensional lattice $\mathbb{Z}^d$ (square/cubic by default, though formulations in other geometries exist \cite{mazzacapo2015}) whereas time remains continuous.
This formulation is particularly well-suited for real-time dynamics and finite chemical potential problems, both of which are challenging for conventional Monte Carlo methods used in Euclidean lattice field theory \cite{troyer_computational_2005}.

The system has local symmetries generated by the gauge group $G$, usually a compact Lie or a finite group as in our case ($\mathbb{D}_8$). These \emph{gauge transformations}, $\Theta_g\left(\mathbf{x}\right)$, are local unitaries under which the physically relevant states and operators are invariant. They are parameterized by group elements  $g \in G$  and are associated with the  sites $\mathbf{x} \in \mathbb{Z}^d$.
The local symmetry introduces conservation laws (\emph{Gauss' law}) on every site. Each $\Theta_g\left(\mathbf{x}\right)$ acts locally on $\mathbf{x}$ and on the links $\ell$ around it (starting or ending at $\mathbf{x}$, see FIG. \ref{Figure1}), transforming only those degrees of freedom in a way that is parametrized by $g$. \\
An operator $A$ is gauge invariant if and only if:
\begin{equation} \label{eqn:gauge_invariance}
 \Theta_g\left(\mathbf{x}\right) A \Theta^{\dagger}_g\left(\mathbf{x}\right) = A, \hspace{30pt} \forall g\in G,\mathbf{x}\in\mathbb{Z}^d;   
\end{equation}
and a gauge-invariant state $\left|\psi\right\rangle$ is  invariant under all gauge transformations (up to a global phase if $G$ is abelian; in the non-Abelian case, gauge transformations can mix the elements of state multiplets  \cite{kasper_from_2020}).

\input{figures/figure1}

 Hence, this formulation allows for the exploration of non-Abelian gauge theories, including QCD, in a more tractable way for real-time simulations.

Despite its advantages, the Hamiltonian formulation of LGTs presents its own challenges. One key difficulty arises from the high dimensionality of the Hilbert space due to the gauge degrees of freedom on each lattice link. Moreover, the inclusion of dynamical fermions introduces further complexities, and others are the need to enforce gauge invariance and generate complicated many-body (plaquette) interactions \cite{zohar_quantum_2022}.

\subsection{The group $\mathbb{D}_8$} \label{D8_group}

Let us focus on the case $G=\mathbb{D}_8$ (dihedral group of order eight or sometimes known as $\mathbb{D}_4$, the symmetry group of a 4-gon) in two space dimensions. This is the group of planar symmetries of a square, which includes $\pi/2$ rotations and reflections. It is a non-Abelian group, generated by a $\pi/2$ rotation, which we denote by $a$, and a diagonal reflection about the origin, denoted by $y$. Labelling the identity element by $e$, we note that
\begin{equation}
a^4=y^2=e.
\end{equation}
Along with
\begin{equation}
    yay=a^{-1}
\end{equation}
one can construct the full multiplication table of the group's eight elements,
\begin{align}
\mathbb{D}_8 &=\left\{e, a, a^2, a^3, y, ay, a^2y, a^3y \right\}\nonumber\\
&=\left\{a^py^q \middle| p=0,1,2,3;\hspace{5pt} q=0,1 \right\}.  
\end{align}


The group has five irreducible representations (irreps). Four are one dimensional unfaithful representations, denoted by $j=0,\overline{0},1,\overline{1}$, representing the group elements as $1,\left(-1\right)^{q},\left(-1\right)^{p},\left(-1\right)^{p+q}$ respectively. This implies that the $1\times 1$ Wigner matrices for these representations are
\begin{equation}
\begin{aligned}
    &D^{0}\left(a\right)=1,\quad D^{0}\left(y\right)=1, \\
    &D^{\Bar{0}}\left(a\right)=1,\quad D^{\Bar{0}}\left(y\right)=-1, \\
    &D^{1}\left(a\right)=-1,\quad D^{1}\left(y\right)=1, \\
    &D^{\Bar{1}}\left(a\right)=-1,\quad D^{\Bar{1}}\left(y\right)=-1.
\end{aligned}
\end{equation}
The only non-trivial and faithful representation, $j=2$, may be constructed (for example) out of the generators' Wigner matrices
\begin{equation}
D^{2}\left(a \right)=-i\sigma_y,\quad D^{2}\left(y\right)=\sigma_z.    
\end{equation}

\subsection{The Hilbert space} \label{D8_Hilbert}

The matter in our model is fermionic. Each site can host two fermionic species of fermions (corresponding to the spinors we have in the nontrivial representation), annihilated by the local Fock operators $\psi_0\left(\mathbf{x}\right)$ and $\psi_1\left(\mathbf{x}\right)$, 
\begin{align}
    \{ \psi_{m}\left(\mathbf{x}\right), \psi_{n}\left(\mathbf{y}\right)\} = 0, \quad
    \{ \psi_{m}\left(\mathbf{x}\right), \psi^{\dagger}_{n}\left(\mathbf{y}\right)\} =
    i\delta_{mn}\delta_{\mathbf{xy}},
\end{align}
where $m,n=0,1$ are group indices within the $j=2$ representation, and $\mathbf{x},\mathbf{y} \in \mathbb{Z}^2$ are lattice sites.
In the following, we assume an implied summation over repeating group indices and, although the discussion can be easily generalized, we focus for concreteness on $d=2$.

\begin{figure}
    \centering
    \includegraphics[width=0.75\linewidth]{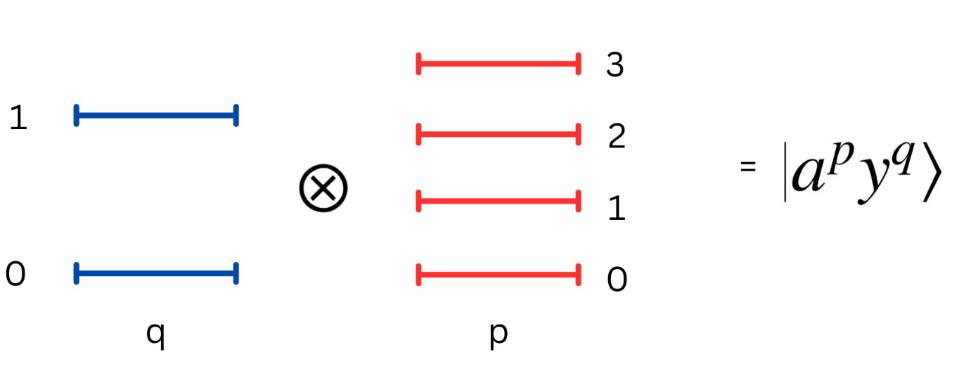}
    \caption{Decomposition of the link hilbert space into a qubit and a qudit.}
    \label{fig:qubit_qudit}
\end{figure}

On the links---which we label by $\left(\mathbf{x}\in\mathbb{Z}^2,i=1,2\right)$, their starting site and the positive direction to which they emanate---the local gauge Hilbert spaces have dimension eight, which equals the group's order. One particular basis of interest is therefore the \emph{group element basis}, $\left\{\ket{g}\right\}_{g\in \mathbb{D}_8}$ where elements are identified with each of the elements of the group. Having the two generators $a,y$, we can represent the elements of this basis in a simpler way, using two quantum numbers, $p=0,1,2,3$ and $q=0,1$:
\begin{equation}
\ket{g} = \ket{a^p y^q} = \ket{p,q}.
\end{equation}
This can be represented exactly by a product  of qubit and  qudit spaces: 
the $q$ quantum number will be encoded by the \emph{qubit} while $p$ by a four-level \emph{qudit}, as can be pictorially seen  in FIG.~\ref{fig:qubit_qudit}.

One can also use the representation basis, whose states $\ket{jmn}$ are labeled by irreps $j$ and eigenvalues of a maximal set of group transformations, $m,n$. The relation between the groups is through the Peter--Weyl theorem, 
\begin{align}
    \braket{g|jmn} = \sqrt{\dfrac{dim(j)}{\abs{G}}} D^j_{mn}(g),
\end{align}
using the Wigner matrices $D^j_{mn}(g)$ \cite{wigner2012group}.
In our case, for the non-trivial representation we have the four states $\ket{2mn}$ where $m,n=0,1$, and the four states of the unfaithful representations, with no $m,n$ indices (since they correspond to singlets), $\ket{0},\ket{\overline{0}},\ket{1},\ket{\overline{1}}$.

To conclude the link description, we define local unitary operators $\theta^R_g$ and $\theta^L_g$ acting on the gauge field and transforming it from the right and left sides, respectively. In the group basis, they act as group multiplication operators \cite{zohar_formulation_2015},
\begin{align}
    &\theta^R_g \ket{h} = \ket{hg^{-1}} \qquad \forall \ h,g \in G,  \label{TR_group basis} \\ 
    &\theta^L_g \ket{h} = \ket{g^{-1}h} \qquad \forall \ h,g \in G, \label{TL_group basis}\\
    &\theta^R_g \ket{jmn} = \ket{jmn'} D^j_{n'n}(g), \\
    &\theta^L_g \ket{jmn} = D^j_{mm'}(g) \ket{jm'n}.
\end{align}
We further introduce the group element operators, $U^j_{mn}$, which will later be used as the \emph{gauge connections}. They are diagonal in the group element basis, satisfying
\begin{equation}
    U^j_{mn} \ket{g} = D^j_{mn}(g) \ket{g}. 
\end{equation}
for every irrep $j$.

The possible gauge transformations are given by
\begin{equation}
\Theta_g\left(\mathbf{x}\right) = W_g\left(\mathbf{x}\right) \theta^\dag_{g,\text{M}}\left(\mathbf{x}\right) \qquad \forall g \in \mathbb{D}_8,
\end{equation}
by which the links and matter sites are transformed as
\begin{align}
\label{operator definitions}
&W_g\left(\mathbf{x}\right) = \prod_{i=1,2} \left[\theta^L_g(\mathbf{x},i) \theta^{R\dag}_g(\mathbf{x} - \hat{\mathbf{e}}_i,i)\right],\\
&\theta^\dag_{g,\text{M}}\left(\mathbf{x}\right) = \text{det}(g^{-1})^{x_1+x_2} e^{-\psi^\dag_a q_{mn}(g)\psi_b}, \\
& q(g) = - i\text{log}(D^2(g)).
\end{align}
The latter is related to the staggered fermions \cite{susskind_lattice_1977} prescription for finite gauge groups, as discussed in Ref.~\cite{zohar_formulation_2015}.
In the above, $W_g(\mathbf{x})$ is a product of operators acting on the links $i$ that are connected to the site $\mathbf{x}$ while $\theta^\dag_{g,\text{M}}$ acts only on the matter. The differentiation between left ($L$) and right ($R$) operators is necessary as $\mathbb{D}_8$ is non-Abelian. $\hat{\mathbf{e}}_i$ are unit vectors in the $i=1,2$ directions,

As we mentioned before, gauge-invariant states $\left|\Psi\right\rangle$ satisfy the local \emph{Gauss' law } constraints, which for $\mathbb{D}_8$, assuming no static charges, can be written in a compact form as
\begin{equation}
	W_g\left(\mathbf{x}\right)\theta^\dag_{g,\text{M}}\left|\psi\right\rangle =  \left|\psi\right\rangle \quad \forall \mathbf{x} \in\mathbb{Z}^2, \forall g \in \mathbb{D}_8.
	\label{eqn:Gauss_D8}
\end{equation} \\

\subsection{The Hamiltonian} \label{D8_Hamiltonian}

The Hamiltonian (see FIG.~\ref{Figure1}) governing the dynamics of the $\mathbb{D}_8$ LGT with dynamical fermions has the form
\begin{equation} \label{eqn:LGT_general_ham}
	H=H_{\text{M}} + H_{\text{E}}+H_{\text{B}}+H_{\text{GM}}.
\end{equation}
For the mass term, $H_{\text{M}}$, we choose to consider the \emph{staggered} formulation \cite{susskind_lattice_1977}. Moreover, in the representation basis, $H_{\text{E}}$ is given by the number of excitations of the gauge field ($j=2$); $n_{\bold{x},m} = \psi^\dag_{\bold{x}, m}\psi_{\bold{x}, m}$, $m=0,1$ (no summation) are the number operators.
Explicitly, the different terms of $H$ are 
\begin{align}
\label{Hamiltonian terms}
    &H_{\text{M}} = M \sum_{\mathbf{x}} (-1)^{x_1+x_2} \psi^{\dagger}_m\left(\mathbf{x}\right)
    \psi_m\left(\mathbf{x}\right),   \\
    &H_{\text{GM}}=\epsilon \sum_{\bold{x}, l=1,2}  \psi^{\dagger}_m\left(\mathbf{x}\right) U_{mn}\left(\bold{x}, l\right) \psi_m\left({\bold{x}+ \hat{\mathbf{e}}_l}\right) + \text{H.c.}  \\
    &H_{\text{B}} = \lambda_B \sum_{p}Tr\big[ U_aU_bU_c^\dag U_d^\dag \big]  \\ 
    &H_{\text{E}} =  \lambda_E \sum_{\mathrm{links}} \ket{2, mn} \bra{2, mn}.
\end{align}
In the above, $H_{\text{E}},H_{\text{B}}$ are the electric and magnetic pure gauge parts, respectively, constituting the Kogut--Susskind Hamiltonian \cite{kogut_hamiltonian_1975} and $H_{GM}$ is the coupling of the gauge fields to the matter \cite{kogut_hamiltonian_1975}. 
The plaquette convention of $H_B$ is given in FIG.~\ref{Figure1}.

While in general LGTs the plaquette term as given in $H_{\text{B}}$ is not Hermitian and requires an addition of the Hermitian conjugate, here it is Hermitian as it is.
The form of $H_{\text{E}}$ for a finite group may be derived using transfer matrix techniques, as in Ref.~\cite{hardware_gonzalez}.

\subsection{Mapping to the qubit--qudit spaces}

To implement a quantum simulation on a qudit platform \cite{ringbauer2022universal}, we need to express the operators involved in the Hamiltonian in terms of gates. If possible, we should map the operators into gates that are easy to implement. 
\noindent Knowing the algebra (Eqns.~\eqref{TR_group basis}, \eqref{TL_group basis}), we can write down the matrix formulation in the group basis and conclude that $\theta_{a^2y}$, $\theta_{y}$, and $\theta_{a^2}$ can be easily translated into a qubit and qudit formulation. Defining $\theta_{a^2y} = \theta_{0}$, $\theta_{y} = \theta_1$, and $\theta_{a^2} = \Pi$ to simplify notation, a possbile mapping to qubit and qudit operators is 
\begin{align}
\label{theta-definitions}
    \theta^L_{a^2y} &=  \theta^L_{0} =X  \otimes \Big (X_{02} + P_{13} \Big ), \\
    \theta^R_{a^2y} &=  \theta^R_{0} =X \otimes \Big (X_{02} + X_{13} \Big ), \notag\\
    \theta^L_{y}    &=   \theta^L_{1} =X \otimes \Big (P_{02} + X_{13} \Big ),  \notag\\
    \theta^R_{y} &=   \theta^R_{1} =X \otimes \bold{1} \notag, \\
    \theta^L_{a^2} &= \theta^R_{a^2} = \Pi =  \bold{1} \otimes \Big (X_{02} + X_{13} \Big ). \notag
\end{align}
In this notation, the first operator acts on the qubit space, while the second operates on the qudit. The operator $Z_{\alpha \beta}$ represents a Pauli-$Z$ operator between the $\alpha$ and $\beta$ levels of the qudit. Similarly, the Pauli-$X$ operator is denoted by $X_{\alpha \beta}$, and the projector into the subspace spanned by the levels $\alpha$ and $\beta$ is given by $P_{\alpha \beta}$. Henceforth, we shall refer to these as the \emph{generalized Pauli gates}. Whenever there are no level indices it is implied that there are only two levels. 

The same mapping to qubit and qudit operators can be done for the connection operators, 
\begin{align}
\label{U-definition}
    &U_{00} = \bold{1} \otimes Z_{02}, \\
    &U_{01} = -Z\otimes Z_{13}, \notag\\
    &U_{10} = \bold{1} \otimes Z_{13}, \notag\\
    &U_{11} = Z \otimes Z_{02}. \notag
\end{align}
In Appendix \hyperref[app_a]{A}, we explain that these are the only operators present in the transformed Hamiltonian after integrating out the fermionic matter.

\section{Elimination of matter}

The transformations we now review  enable physics simulation on platforms that do not naturally support fermionic degrees of freedom (e.g., trapped ions or superconducting circuits). In this way, we expand the potential available hardware for quantum simulating LGTs with dynamical matter beyond ultra-cold atomic systems, the original candidates \cite{zohar_quantum_2016}. 

This transformation has a second advantage. The dynamics of LGTs is highly constrained due to the local symmetries on every site. As a result, out of the full Hilbert space of matter and gauge fields, only an exponentially small part is physical and needs to be simulated.
We resolve this redundancy by integrating out the matter, which significantly decreases the overhead for implementation on the quantum hardware. 
The final result will be a simplified problem where we only deal with gauge degrees of freedom without losing the properties of fermionic statistics. 

\subsection{Bosonic reformulation of the matter} \label{method1}
Following the procedure of Ref.~\cite{zohar_eliminating_2018}, we can replace the fermionic matter of any LGT whose gauge group $G$ contains a group element whose non-trivial representation is $-\bold{1}$. For concreteness, we apply this to the smallest possible non-Abelian  group; while $\mathbb{D}_6$ is the smallest non-Abelian group, it does not contain such an element, and thus we opt for $\mathbb{D}_8$. 

In this procedure, one applies a unitary and local transformation $\mathcal{V}^{\text{(1)}}$ to a product of a state of the original LGT, $\ket{\psi}$, and trivial states of some auxiliary degrees of freedom  (which can be disregarded at the end) and employs local, stringless  Jordan--Wigner transformations. The result is a state $\left|\psi^{(1)}\right\rangle$ where the matter is  hard-core bosonic instead of fermionic, and the interaction ranges, while slightly extended, are still local \cite{zohar_eliminating_2018}. This is possible thanks to the local symmetry and in particular its $\mathbb{Z}_2$ central subgroup, which allows one to absorb the statistics by the gauge fields, instead of by the matter.

The hardcore bosonic matter is annihilated by the operators $\eta_m\left(\mathbf{x}\right)$, with $m=0,1$ referring to the two species of matter. Hardcore bosonic operators on different sites commute, i.e.,
\begin{equation}
    \left[\eta_m\left(\mathbf{x}\right),\eta_n\left(\mathbf{y}\right)\right]=
    \left[\eta_m\left(\mathbf{x}\right),\eta^{\dagger}_n\left(\mathbf{y}\right)\right]=0,
    \quad \forall \mathbf{x}\neq\mathbf{y},
\end{equation}
but anti-commute on-site,
\begin{equation}
    \left\{\eta_m\left(\mathbf{x}\right),\eta_n\left(\mathbf{x}\right)\right\}=0,\quad
    \left\{\eta_m\left(\mathbf{x}\right),\eta^{\dagger}_n\left(\mathbf{x}\right)\right\}=\delta_{mn}.
\end{equation}

The Hamiltonian is transformed according to $\mathcal{V}^{\text{(1)}}$, and while the electric part is left unchanged, the other parts obtain the form
 \begin{equation}\label{hamiltonian terms after 1st trasf}
 \begin{aligned}
      &H^{(1)}_{\text{M}} = M \sum_{\mathbf{x}} (-1)^{x_1+x_2} \eta^{\dagger}_m\left(\mathbf{x}\right)
    \eta_m\left(\mathbf{x}\right),   \\
     &H^{(1)}_{\text{B}} = \lambda_B \sum_{p} \xi_p Tr\big[ U_aU_bU_c^\dag U_d^\dag \big], \\
     &H^{(1)}_{\text{GM}} 
     =-i\epsilon \sum_{\bold{x}, l=1,2}\xi_l(\bold{x})  \eta^{\dagger}_m\left(\mathbf{x}\right) U_{mn}\left(\bold{x}, l\right) \eta_m\left({\bold{x}+ \hat{\mathbf{e}}_l}\right) + \text{H.c.} .
    \end{aligned}
    \end{equation}
Here, the statistics is taken care of by the (operatorial) phase factors \cite{zohar_formulation_2015}
\begin{equation}
\begin{aligned}
&  \xi_h(\bold{x}) = \Pi_v(x) \Pi_h(\bold{x- e_1}) \Pi_v(\bold{x - e_2}) \Pi_v(\bold{x+ e_1 - e_2}) \\ 
  & \xi_v(\bold{x}) = \Pi_h(\bold{x - e_1}) \Pi_v(\bold{x - e_2}) \\
  & \xi_p = \Pi_h(\bold{x}) \Pi_v(\bold{x + e_1}) \Pi_v(\bold{x + e_2}) \Pi_h( \bold{x- e_1 + e_2}),
\end{aligned}
\end{equation}
where the $h,v$ subscripts refer to links emanating from the respective sites horizontally and vertically, respectively.  

In the Gauss' laws \ref{eqn:Gauss_D8}, the fermionic operators are simply replaced by the hard-core bosonic ones.

\subsection{Matter removal} \label{method2}
\noindent  To reduce the required amount of quantum resources, and to circumvent the redundancy problem mentioned above, one approach is to eliminate the gauge fields by fully integrating them out. However, this is only feasible in the one dimensional case (see, e.g., \cite{hamer_lattice_1979, martinez_real-time_2016,banuls_efficient_2017, sala_variational_2018}). In higher dimensions, this procedure is not achievable, but an alternative strategy is to eliminate the bosonic matter instead. In Ref.~\cite{zohar_removing_2019},  an elimination scheme for the matter in U$(N)$ and SU$(N)$ theories was proposed.
The main idea is to build a unitary transformation, $\mathcal{V}^{(2)}$, which will decouple the matter degrees of freedom from the gauge fields, leaving us with a product of a gauge field state containing all the information, and a trivial matter state that can be disregarded.

Here, we proceed to a complete elimination of the matter for a $\mathbb{D}_8$ LGT. In this case, on each site, we have eight Gauss' laws, or constraints, corresponding to the eight group elements; however, only two are necessary to eliminate the two species of matter we have. From the explicit form (Eq.~\eqref{eqn:Gauss_D8}), we observe that the laws associated with the $a^2y$ and $y$ elements result in two independent  Gauss' laws, corresponding to two commuting transformations. These provide us with separate, independent constraints for each matter component: 
\begin{align}
\label{Gauss laws for D8}
    & W_{y}(\mathbf{x}) \ket{\psi^{\paren{1}}} = \varepsilon(\mathbf{x}) \big(1 - 2n_1(\mathbf{x}) \big )\ket{\psi^{\paren{1}}} , \\ 
    & W_{a^2y} (\mathbf{x})  \ket{\psi^{\paren{1}}} = \varepsilon(\mathbf{x}) \big(1 - 2n_0(\mathbf{x}) \big )\ket{\psi^{\paren{1}}} , 
\end{align}
where, for notation simplicity, we define the staggering term as $(-1)^{x_1+x_2} \equiv \varepsilon(\mathbf{x})$; we also use the number operators $n_1(\mathbf{x}) = \eta^\dag_1(\mathbf{x}) \eta_1(\mathbf{x})$ and $n_1(\mathbf{x}) = \eta^\dag_0(\mathbf{x}) \eta_0(\mathbf{x})$. Introducing the projectors
\begin{align}
    &P_1^\pm(\bold{x}) = \dfrac{1}{2} \Bigg [ 1 \mp \varepsilon(\bold{x}) W_1(\bold{x})\Bigg], \\ 
    &P_0^\pm(\bold{\bold{x}}) = \dfrac{1}{2} \Bigg [ 1 \mp \varepsilon(\bold{x}) W_0(\bold{x})\Bigg] ,
    \label{projectors}
\end{align}
the two independent Gauss' laws can be written as 
\begin{align}
\label{eqn:gauss_laws_with_number_operators}
    & P_0^+(\bold{x}) \ket{\psi^{\paren{1}}} = n_0(\bold{x}) \ket{\psi^{\paren{1}}}  \qquad \forall \bold{x},\\
    & P_1^+(\bold{x}) \ket{\psi^{\paren{1}}}  = n_1(\bold{x}) \ket{\psi^{\paren{1}}}  \qquad \forall \bold{x}.
    \label{eqn:gauss_laws_with_numeber_2}
\end{align}
Finally, we apply a local Jordan--Wigner transformation \cite{jordan_uber_1928} on each site $\mathbf{x}$,
\begin{align}
\label{local JW}
    \eta_0(\bold{x}) \longrightarrow \tau_0^+(\bold{x}), \qquad \eta_1(\bold{x}) \longrightarrow - \tau_0^z(\bold{x})\tau_1^+(\bold{x}),
\end{align}
where the $\tau$ are Pauli operators.
We can then reformulate Eq.~\eqref{eqn:gauss_laws_with_number_operators} and Eq.~\eqref{eqn:gauss_laws_with_numeber_2} in a similar form:
\begin{align}
\label{eqn:Gauss_with_projectors}
&\big(P_1^{+}(\bold{x})-P_1^{-}(\bold{x})\big)\left|\psi^{(1)}\right\rangle = \tau^z_1(\bold{x})\left|\psi^{(1)}\right\rangle, \\
&\big(P_0^{+}(\bold{x})-P_0^{-}(\bold{x})\big)\left|\psi^{(1)}\right\rangle = \tau^z_0(\bold{x})\left|\psi^{(1)}\right\rangle. 
\end{align}
Using these constraints, we can write down a couple of unitaries that locally eliminate the two types of matter we have: 
\begin{align}
   &\mathcal{V}_1(\bold{x}) = P_1^+(\bold{x}) \otimes \tau_1^{\text{x}}(\bold{x}) + P_1^-(\bold{x}) \otimes \bold{1},\\ 
   &\mathcal{V}_0(\bold{x}) = P_0^+(\bold{x}) \otimes \tau_0^{\text{x}}(\bold{x}) + P_0^- (\bold{x})\otimes \bold{1}.
   \label{Guy Unitary}
\end{align}
The rationale behind this transformation is as follows. If the gauge fields connected to a vertex are projected onto the $+$ subspace, the Gauss' law implies there is a charge at the site, which is removed by the $\tau^{\text{x}}$ operator. If they are projected onto $-$, no charge is present. In both cases, the final charge state is the vacuum. 

Since $\left[\mathcal{V}_n(\bold x),\mathcal{V}_m(\bold{y})\right]=0$, we can safely define
\begin{equation}
\label{eqn:second_transf}
    \mathcal{V}^{(2)} = \underset{\bold{x}}{\prod}\mathcal{V}_1(\bold{x})\mathcal{V}_0(\bold{x}).
\end{equation}

$\mathcal{V}^{\paren{2}}$ is unitary and it follows that the spectrum of the physical problem does not change even though we have a different Hamiltonian $H^{(2)} = \mathcal{V}^{(2)} H \mathcal{V}^{(2)\dag}$. 
We label the transformed states as follows: 
\begin{align}
    \left|\psi^{(2)}\right\rangle&= \mathcal{V}^{(2)}\left|\psi^{(1)}\right\rangle.
    \end{align}
After applying $\mathcal{V}^{(2)}$, the matter at all sites now resides in the ground state (apply $\mathcal{V}^{\paren{2}}$ to \fulleqref{eqn:Gauss_with_projectors}):
\begin{equation} \label{eqn:gauss_law_method2}
	\tau^z_n(\mathbf{x})\left|\psi^{(2)}\right\rangle=-\left|\psi^{(2)}\right\rangle, \hspace{10pt} \forall \mathbf{x}, \qquad n=0,1.
\end{equation}
Consequently, the matter spaces can be projected out by working with 
\begin{align}
    \Tilde{H}^{(2)} = \braket{\Omega | H^{(2)}  | \Omega}, \qquad  \text{where }\ket{\Omega} = \prod_\bold{x} \ket{\downarrow}_{\bold{x},1}\ket{\downarrow}_{\bold{x},0}.
\end{align}
This procedure (applying $\mathcal{V}^{(2)}$ to $H^\text{(1)}$ and projecting on $\ket{\Omega}$) is described with more details in Appendix \hyperref[app_a]{A}. Note that $H_E$, $H_M$, and $H_B$ are invariant under $\mathcal{V}^{(2)}$.
With these transformations, the entire LGT including staggered fermions is encoded into a local Hamiltonian acting only on the gauge degrees of freedom.

%% file: figures/figure1.tex
\begin{figure}
  \centering
    \begin{tikzpicture}
  \foreach \x in {1, 2, 3, 4}
    \foreach \y in {2, 3, 4}
      \fill (\x,\y) circle (2pt); 
      
\foreach \x in {0.5, 1.5, 2.5, 3.5}
    \foreach \y in {1.5, 2.5, 3.5}
      \fill (\x,\y) circle (2pt); 

  \foreach \x in {0.5, 1.5, 2.5, 3.5}
    \foreach \y in {2, 3, 4}
        \draw(\x-0.05,\y-0.05) rectangle (\x+0.05,\y+0.05);
     \node[blue] at (3,2.7) {$U_a$};
     \node[blue] at (3.8,3.5) {$U_b$};
     \node[blue] at (3,4.3) {$U^\dag_c$};
     \node[blue] at (2.2,3.5) {$U^\dag_d$};
     \node[blue] at (3,3.5) {$p$};
     \fill[blue] (3,4) circle (2pt);
     \fill[blue] (3,3) circle (2pt);
     \fill[blue] (3.5,3.5) circle (2pt);
     \fill[blue] (2.5,3.5) circle (2pt);
     \node[green!65!black] at (0.5,1.75) {$M$};
     \fill[red] (4,2) circle (2pt);
     \draw[blue] (2.5,3) rectangle (3.5,4);
     \fill[purple] (1,3) circle (2pt);
     \draw[purple, fill=purple] (1.5-0.05,3-0.05) rectangle (1.5+0.05,3+0.05);
     \draw[purple, fill=purple] (0.5-0.05,3-0.05) rectangle (0.5+0.05,3+0.05);
     \draw[green!65!black, fill=green!65!black] (0.5-0.1,2-0.05) rectangle (0.5+0.05,2+0.05);
     \node[purple] at (0.5,2.7) {$\psi^\dag$};
     \node[purple] at (1.5,2.7) {$\psi$};
     \node[purple] at (1,2.7) {$U$};
     
     \draw[orange, fill=orange] (2.5-0.05,2-0.05) rectangle (2.5+0.05,2+0.05);
     \fill[orange] (3,2) circle (2pt);
     \fill[orange] (2.5,2.5) circle (2pt);
     \fill[orange] (2.5,1.5) circle (2pt);
     \fill[orange] (2,2) circle (2pt);
    \draw[orange] (2.5,2) -- (3,2);         
    \draw[orange] (2.5,2) -- (2.5,2.5); 
    \draw[orange] (2.5,2) -- (2.5,1.5); 
    \draw[orange] (2.5,2) -- (2,2);     
\end{tikzpicture}
    \caption{Illustration of gauge-invariant operators in the $d=2$ Hamiltonian formulation of LGT. Squares represent matter sites while circles are the gauge fields on the links. The orange cross depicts a local gauge transformation as described in Eq.~\eqref{eqn:Gauss_D8}. The red circle is an example of the electric Hamiltonian term we have on all the links; this come from $H_E$. The blue plaquette stands for the magnetic Hamiltonian term $H_B$ and the green square is an example of the local mass term $H_M$. The purple horizontal line represents a hopping interaction $H_{GM}$.}
    \label{Figure1}
\end{figure}

%% file: figures/1d_figure.tex
\begin{figure}[!h]
    \centering
    \begin{tikzpicture}
  \fill (-2,0) circle (3pt);
\fill (-4,0) circle (3pt);
  \fill (0,0) circle (3pt);
  \fill (2,0) circle (3pt);
  \fill[blue] (-3,0) circle (3pt); 
\fill[blue] (-1,0) circle (3pt); 
\fill[blue] (1,0) circle (3pt); 
 
  \node at (2,-0.4) {$x + e_1$}; 
  \node at (0, -0.4) {$x$}; 
  \node at (-2,-0.4) {$x - e_1$};
\node at (-4,-0.4) {$x - 2e_1$};


    \node[blue]  at (-3, -0.4) {$l$}; 
  \node[blue]  at (-1, -0.4) {$c$}; 
  \node[blue] at (1,-0.4) {$r$};
\end{tikzpicture}

\begin{tikzpicture}

\draw[->, thick] (-1,1.3) -- (-1,0.3);
  \fill[blue] (-3,0) circle (3pt); 
\fill[blue] (-1,0) circle (3pt); 
\fill[blue] (1,0) circle (3pt); 

\node at (0,0.8) {$\mathcal{V}^{(2)} \mathcal{V}^{(1)} $};
  \node[blue]  at (-3, -0.4) {$l$}; 
  \node[blue]  at (-1, -0.4) {$c$}; 
  \node[blue] at (1,-0.4) {$r$};
\end{tikzpicture}
\caption{1D model, on the top the original model, where the gauge field is represented in blue, and the matter field in black. The lower part shows the transformed model, now featuring only the links, denoted as $c$, $l$, and $r$, corresponding to the center, left, and right, respectively. These labels will be used in subsequent references to indicate the specific link under consideration.}
    \label{1d_model_picture}
\end{figure}

%% file: Sections/1d_analysis.tex
We begin by analyzing a simple one-dimensional system. This approach allows us to examine the various components of the Hamiltonian, including their representation in terms of gates. We also perform benchmark simulations, whose results align well with theoretical expectations.
\subsection{Theory}\label{1d_theory}
The one-dimensional transformed interaction Hamiltonian can be obtained from the general formulation as a special case. As there are no closed loops, we do not have the magnetic (plaquette) term. We use a link-only formulation, where the matter sites are fully eliminated  using the procedure given above (see FIG.~\ref{1d_model_picture}). 

In the gate formulation (Eqns.~\eqref{theta-definitions},\eqref{U-definition}), the interaction Hamiltonian is given by
\begin{equation}
\tilde{H}_{\text{GM}}=\frac{J}{2}\sum_{\mathbf{x}}\sum_{mn}\hat{H}_{mn}\left(\mathbf{x}\right).
\label{summation}
\end{equation}
For each link $\mathbf{x}$, the local interaction term  consists of four terms,  involving three links---the link $\mathbf{x}$ and its nearest neighbors. To ease to notation, we denote the middle link by $c$ and its neighbors by $l$ and $r$, see FIG.~\ref{1d_model_picture}. Then, the interaction terms take the form
\begin{align}
\hat{H}_{00} & =\left(X_{02}^{l}+X_{13}^{l}\right)X^{c}Y_{02}^{c}X^{r}\left(X_{02}^{r}+P_{13}^{r}\right)+X^{l}X^{c}Y_{02}^{c} \label{eq:H_int_qubit_final} \\
\hat{H}_{01} & =-\left(X_{02}^{l}+X_{13}^{l}\right)Z^{c}Y_{13}^{c}\left(X_{02}^{r}+X_{13}^{r}\right)-X^{l}Z^{c}Y_{13}^{c}X^{r}\left(X_{02}^{r}+P_{13}^{r}\right)\nonumber \\
\hat{H}_{10} & =-Y_{13}^{c}-X^{l}Y_{13}^{c}X^{r}\left(X_{02}^{r}+P_{13}^{r}\right) \nonumber\\
\hat{H}_{11} & =-X^{l}Y^{c}Z_{02}^{c}\left(X_{02}^{r}+X_{13}^{r}\right)-Y^{c}Z_{02}^{c}X^{r}\left(X_{02}^{r}+P_{13}^{r}\right).\nonumber 
\end{align}

A sum of such terms will give the interaction Hamiltonian for a closed chain. We run our simulation with a chain of four sites and open boundary conditions, hence resulting in only three links for which we can simply keep the $l,c,r$ convention. The middle link will give rise to contributions as above, but we also need to include boundary terms from the transformed contributions of the interaction Hamiltonian on the $l,r$ links.
For the boundary links (left, L ,and right, R) we get:
\begin{align}
\hat{H}_{00}^{\text{L}} & =-X^{l}Y_{02}^{l}X^{c}\left(X_{02}^{c}+P_{13}^{c}\right)-X^{l}Y_{02}^{l} \label{eq:H_int_left_boundary} \\
\hat{H}_{01}^{\text{L}} & =-Z^{l}Y_{13}^{l}\left(X_{02}^{c}+X_{13}^{c}\right)-Z^{l}Y_{13}^{l}X^{c}\left(X_{02}^{c}+P_{13}^{c}\right)\nonumber \\
\hat{H}_{10}^{\text{L}} & =-Y_{13}^{l}-Y_{13}^{l}X^{c}\left(X_{02}^{c}+P_{13}^{c}\right)\nonumber \\
\hat{H}_{11}^{\text{L}} & =+Y^{l}Z_{02}^{l}\left(X_{02}^{c}+X_{13}^{c}\right)+Y^{l}Z_{02}^{l}X^{c}\left(X_{02}^{c}+P_{13}^{c}\right) \nonumber \\
\hat{H}_{00}^{\text{R}} & =-\left(X_{02}^{c}+X_{13}^{c}\right)X^{r}Y_{02}^{r}-X^{c}X^{r}Y_{02}^{r} \label{eq:H_int_boundary_R}\\
\hat{H}_{01}^{\text{R}} & =-\left(X_{02}^{c}+X_{13}^{c}\right)Z^{r}Y_{13}^{r}-X^{c}Z^{r}Y_{13}^{r}\nonumber 
\end{align}
\begin{align}
\hat{H}_{10}^{\text{R}} & =-Y_{13}^{r}-X^{c}Y_{13}^{r} \nonumber\\
\hat{H}_{11}^{\text{R}} & =+X^{c}Y^{r}Z_{02}^{r}+Y^{r}Z_{02}^{r}.\nonumber 
\end{align}

The full interaction term is given by
\begin{equation}
\label{interaction_term_1d}
    \tilde{H}_{\text{GM}}=\frac{J}{2}\sum_{mn}\left(\hat{H}_{mn}^{\text{L}}+\hat{H}_{mn}+\hat{H}_{mn}^{\text{R}}\right),
\end{equation}

\noindent The electric field Hamiltonian and the mass term, in the qubit--qudit scheme, are given by 
\begin{align}
H_{\text{E}}&=-\frac{\lambda_{E}}{2}\sum_{i= l,c,r}\left(X^i_{02}+X^i_{13}\right), \label{eletric_1D}
\end{align}
\begin{align}
H_{\text{M}}&= -\dfrac{M}{2} \sum_{i= 1,...,4} \hat{H}_{\text{Mi}},
\label{mass_1d_form}
\end{align} 
where the $i$ index stands for the different vertices we are considering. The different terms are given by 
\begin{align}
\hat{H}_{\text{M}1}& =X^{l}\left(X_{02}^{l}+P_{13}^{l}\right) + X^{l}\left(P_{02}^{l}+X_{13}^{l}\right)  \label{eq:H_m_in_gates} \\
\hat{H}_{\text{M}2} &=X^{l}\left(X_{02}^{l}+X_{13}^{l}\right)X^{c}\left(X_{02}^{c}+P_{13}^{c}\right) + X^{l}X^{c}\left(P_{02}^{c}+X_{13}^{c}\right) \nonumber \\
\hat{H}_{\text{M3}} &=X^{c}\left(X_{02}^{c}+X_{13}^{c}\right)X^{r}\left(X_{02}^{r}+P_{13}^{r}\right) + X^{c}X^{r}\left(P_{02}^{r}+X_{13}^{r}\right) \nonumber \\ 
\hat{H}_{\text{M}4}& =X^{r} + X^{r}\left(X_{02}^{r}+P_{13}^{r}\right) \nonumber
\end{align}

Finally, the Hamiltonian we want to simulate for three links is
\begin{align}
H= H_{\text{E}} + H_{\text{M}} + H_{\text{GM}},
\label{eq:hamiltonian_1D_model}
\end{align}
where $H_{\text{GM}},H_{\text{E}},H_{\text{M}}$
are given in Eqns.~\eqref{interaction_term_1d} ,\eqref{eletric_1D},
and \eqref{mass_1d_form}.

\subsection{Simulations}
\label{sec:simulations_1D}

Here, we present the numerical results for the variational ground-state preparation of the one-dimensional three-link system. We focus on the challenging point of the phase diagram $J = M = h = 1$, where all terms in the Hamiltonian compete.

The PQC we use in this case is given in Fig.~\ref{fig:circuit_1D}. 
It is composed of single-qudit and single-qubit operations, as well as entangling two-qudit and qubit--qudit operations. The number of entangling operations in each layer is chosen to be $L_\mathrm{ent} = L_b + L_d$, where $L_{b/d}$ is the number of qubits and qudits in the system, respectively. Therefore, for a circuit of $N$ layers, the total number of entangling operations that need to be implemented is $L_\mathrm{ent} = N(L_b + L_d)$.

We employ the VarQITE algorithm, starting from the initial product state $\ket{\psi_0} = \bigotimes^3_{i = 1}\ket{1}_{b,i}\otimes\ket{3}_{d,i}$ and initializing the variational parameters of the circuit as $\boldsymbol{\theta} = \boldsymbol{0}$. The equation of motion with respect to the imaginary time for the variational parameters (Eq.~\eqref{eq:eom_for_the_thetas}) is solved numerically, producing the trajectories $\boldsymbol{\theta}(\tau)$, which mimic the imaginary time evolution w.r.t. the Hamiltonian $H$ of Eq.~\eqref{eq:hamiltonian_1D_model}. In order to take into account the gauge symmetry of the model, we further include a penalty term for the transformed Gauss' law constraint (that is, after the elimination of fermions); for more details we refer to Section \ref{sec:solution_of_pure_gauge}. This is done by adding to the Hamiltonian of the system an extra term $H \rightarrow H + \lambda\sum_i G_i^2$, where $G_i$ is the Gauss' law operator at site $i$ and $\lambda$ is a Lagrange multiplier. For each time $\tau$ and set of parameters $\boldsymbol{\theta}(\tau)$, we calculate the expectation value of the Hamiltonian in the trial state $E(\boldsymbol{\theta}) = \bra{\psi(\boldsymbol{\theta})}H\ket{\psi(\boldsymbol{\theta})}$ as well as the fidelity w.r.t. the ground state $\ket{\psi_{\text{ground}}}$ of $H$ obtained by exact diagonalization, defined by
\begin{align}
    \mathcal{F}_{\mathbf{\theta}}(\tau) = |\langle\psi(\mathbf{\theta}(\tau))|\psi_{\text{ground}}\rangle|^2
\label{eq:fidelity}
\end{align}

In FIG.~\ref{fig:1d_energy} and FIG.~\ref{fig:1d_fidelity}, those two quantities are shown as a function of the imaginary time $\tau$. For a variational circuit, composed of a total number of $N = 3$ layers, the fidelity with respect to the ground state is already as high as $99\%$. The total number of entangling operations in the circuit with $N = 3$ layers is $L_\mathrm{ent} = 3(3+3) = 18$.

%% file: Sections/2D_Case.tex
In this section, we present the theoretical analysis of a pseudo-2D system which we treat through an exact analytical method. This allows us to demonstrate the effectiveness of the transformations in the non-Abelian case. We provide also the theoretical analysis of a plaquette, the smallest non-trivial 2D-system. 

Above, we have analyzed the transformation properties of the Hamiltonian under various transformations \cite{zohar_eliminating_2018, zohar_removing_2019} and provided a simple one dimensional example. Here, we generalize to higher dimensions, where these transformations are really needed, and identify the complexity of the interactions in the Hamiltonian in terms of generalized Pauli gates.
The system involves many-body interactions; for example, the horizontal interaction Hamiltonian contains a five-link and a seven-link interaction. Importantly, in our formulation, each link comprises both a qubit and a qudit, meaning a seven-link interaction could involve more than seven bodies. Table \ref{tab_hamiltonian} provides a concise summary of the interactions.

Given these considerations, the terms may involve up to interacting 12 bodies, rendering it impractical to implement this theory with current hardware capabilities, either through a purely analog or digital approach. Alternatively, we can employ a hybrid quantum-classical protocol for quantum simulation--the variational time-evolution algorithm. In this approach, the experimental overhead one is willing to spend can be directly chosen, and the accuracy of the algorithm can be estimated accordingly. 

\begin{table}[h]
    \centering
    \begin{tabular}{c|c|c|}
        & Max $\#$ bodies & structure of the ``hardest" term \\
        \hline 
         Electric term      & 1 & 1 qudit\\
         Magnetic term      & 10& 4 qubits + 6 qudits\\
         Vertical int.      & 12& 7 qubits + 5 qudits\\
         Horizontal int.    & 11& 7 qubits + 4 qudits\\ 
         \end{tabular}
    \caption{Features of the various components of the Hamiltonian. By ``hardest" we refer to the component involving the highest number of bodies. From an experimental standpoint, the notion of hardness may vary, given that qubits and qudits present differing implementation challenges.}
    \label{tab_hamiltonian}
\end{table}

\subsection{Analytical verification of matter elimination: pseudo-2D}\label{2d_pseudo_theory}
\input{figures/figure3}

In this section, we aim to verify the correctness of the transformations in the non-Abelian case and the resulting Hamiltonian. The most effective approach is to ensure that the spectrum of the Hamiltonian with matter aligns with that of the transformed one. While the primary objective of quantum simulation is to find spectra and states of complex systems, it is imperative to validate the model's accuracy.
It may seem surprising that fermionic matter and the substantial redundancy can be addressed in this manner within a non-Abelian LGT. However, we demonstrate analytically that this approach is consistent and free of any irregularities.

A pseudo-2D model (see FIG.~\ref{Figure3}) represents the simplest non-trivial 2D lattice available to benchmark the results obtained in the preceding section. In this simple model, we are foregoing the inclusion of the magnetic term, but alternative techniques can still be employed to validate the complete model as we did in \ref{Magnetic_verification}.

Our strategy is as follows. 
First, we identify the gauge-invariant states and corresponding spectrum in the presence of matter. Next, we apply the transformations to these states and evaluate the spectrum. If the transformation is correct, the spectra must coincide. Subsequently, we derive the states from a pure gauge theory with the transformed constraints and verify that they match the original ones. Once this is established, we demonstrate that the transformations hold in the non-Abelian case. 

\subsubsection{Solution of the original model}
We impose gauge invariance and determine the states satisfying the Gauss' laws. Drawing an analogy with the Wigner--Eckart theorem \cite{wigner2012group}, we seek to ensure that, for each site, the combination of right irreps equals the sum of left irreps. This can be achieved by employing Clebsch--Gordan (CG) coefficients and subsequently verifying that the state maintains gauge invariance.
The most general gauge-invariant state is given by (for the notation refer to FIG.~\ref{Figure3})
\begin{align}
    \ket{\Psi} = &\sum_{J_i, M_j, N_k} \alpha^{J_1J_2J_3j_aj_bj_cj_d}_{M_1M_2M_3N_1N_2N_3} \ket{J_1M_1N_1}\ket{J_2M_2N_2} \ket{J_3M_3N_3} \times \notag \\ 
    & \times (a_0^\dag)^{n_{a_0}} (a_1^\dag)^{n_{a_1}} (b_0^\dag)^{n_{b_0}} (b_1^\dag)^{n_{b_1}} (c_0^\dag)^{n_{c_0}} (c_1^\dag)^{n_{c_1}} (d_0^\dag)^{n_{d_0}} (d_1^\dag)^{n_{d_1}} \ket{\Omega},
\end{align}
where $\ket{\Omega}$ stands for the matter \textit{Fock vacuum}. In addition, we impose the half-filling condition for the fermions.

\begin{figure}
    \centering
    \includegraphics[scale = 0.55]{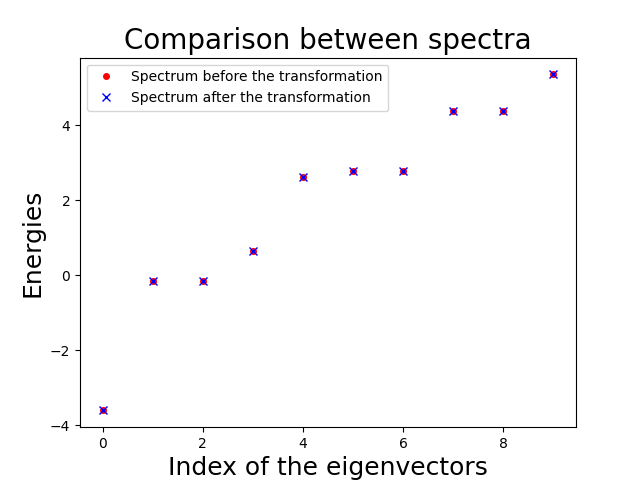}
    \caption{Spectrum evaluated with $M = 1$, $\lambda_E = 1$, and $J=1$. The solutions of the original and transformed model yield the same result, confirming the validity of the employed transformations.}
    \label{Spectrum_comparison}
\end{figure}

By utilizing the Clebsch--Gordan coefficients detailed in Table \ref{C-G coeff}, we find a total of ten gauge-invariant states, $\{ \ket{\Psi_i}\}_{i = 1,...10}$, for which we checked that they respect the different Gauss' laws. Moreover, we also assessed their rotational symmetry (under a rotation every state is mapped into another one) and verified that the Hamiltonian is rotationally invariant. The spectrum of the ten states is given in FIG.~\ref{Spectrum_comparison}.

\begin{table}[h]
    \centering
    \begin{tabular}{c|c|}
         coefficients &  value\\
         $ \braket{0|00}, \qquad \braket{0|\Bar{0}\Bar{0}}, \qquad \braket{0|11}, \qquad \braket{0|\Bar{1}\Bar{1}}$    &   $1$ \\ 
    $ \braket{\Bar{0}|\Bar{0}0}, \qquad \braket{\Bar{0}|0\Bar{0}}, \qquad \braket{\Bar{0}|1\Bar{1}}, \qquad \braket{\Bar{0}|\Bar{1}1}$    &   $1$ \\ 
    $ \braket{1|10}, \qquad \braket{1|01}, \qquad \braket{1|\Bar{0}\Bar{1}}, \qquad \braket{1|\Bar{1}\Bar{0}}$    &   $1$ \\
    $ \braket{\Bar{1}|\Bar{1}0}, \qquad \braket{\Bar{1}|0\Bar{1}}, \qquad \braket{\Bar{1}|\Bar{0}1}, \qquad \braket{\Bar{1}|1\Bar{0}}$ & 1\\ 
    $ \braket{0,2m|2m'}, \qquad \braket{2m, 0|2m'}$ & $\delta_{m,m'}$\\ 
    $ \braket{\Bar{0},2m|2m'}, \qquad \braket{2m, \Bar{0}|2m'}$ & $\varepsilon_{m,m'}$\\ 
    $ \braket{1,2m|2m'}, \qquad \braket{2m, 1|2m'}$ & $(\sigma_z)_{m,m'}$\\ 
    $ \braket{\Bar{1},2m|2m'}, \qquad \braket{2m, \Bar{1}|2m'}$ & $(\sigma_x)_{m,m'}$\\ 
    $ \braket{0|2m, 2m'} $ & $\frac{1}{\sqrt{2}}\delta_{m,m'}$\\
    $ \braket{\Bar{0}|2m, 2m'} $ & $\frac{1}{\sqrt{2}}\epsilon_{m,m'}$\\
    $ \braket{1|2m, 2m'} $ & $\frac{1}{\sqrt{2}}\left(\sigma_z\right)_{m,m'}$\\
    $ \braket{\Bar{1}|2m, 2m'} $ & $\frac{1}{\sqrt{2}}\left(\sigma_x\right)_{m,m'}$\\
    \end{tabular}
    \caption{Clebsch--Gordan coefficients, $m, m' \in \{0,1\}$, computed using the Clebsch--Gordan series \cite{brink1994angular}: $D^{j_1}_{m_1m_1'} D^{j_2}_{m_2m_2'} = \sum_J \braket{j_1m_1j_2m_2|JM} \braket{JM'|j_1m_1'j_2m_2'}D^J_{MM'}$.}
\label{C-G coeff}
\end{table}

\subsubsection{Solution of the transformed model}
\label{solution of the trasformed model}

In the preceding section, we have identified the ten gauge-invariant states, evaluated the Hamiltonian upon them, and plotted the resulting spectrum. Now, we aim to determine the spectrum of the pseudo-2D model in the absence of matter and contrast it with the original one. We apply $\mathcal{V}^{(2)}$ (Eq.~\eqref{eqn:second_transf}) to the Hamiltonian and to the gauge invariant states found before. For the Hamiltonian, the process is similar to the 2D analysis, see Appendix~\hyperref[app_a]{A} for more details. 
In order to validate the transformed Hamiltonian, we proceed to compute the gauge-invariant states under $\mathcal{V}^{(2)}$. This computation allows us to compare the resulting spectrum with the analysis that includes the matter (FIG.~\ref{Spectrum_comparison}). Moreover, from the explicit calculation we found that, as expected, the states $\{ \ket{\psi^{(2)}_i}\}_{i=1,2...10}$ are simply deprived of the matter part, while the gauge fields remain unchanged. \\

\subsubsection{Solution of a pure gauge theory}
\label{sec:solution_of_pure_gauge}
It is essential to verify that the ten states are exhaustive. To achieve this, we begin with a pure-gauge theory and impose Gauss' laws. Since matter is no longer present, we must derive a new version of the Gauss' laws, which we achive by applying $\mathcal{V}^{(2)}$ to Eq.~\eqref{eqn:Gauss_D8}:
\begin{equation}
	    W^{(2)}_g\left(\mathbf{x}\right)\theta^{\dag(2)}_{g,\text{M}}\left(\mathbf{x}\right) \left|\psi^{(2)}\right\rangle =  \left|\psi^{(2)}\right\rangle \quad \forall \mathbf{x} \in\mathbb{Z}^d, \forall g \in \mathbb{D}_8.
	\label{eqn:Gauss_D8_trasformed}
\end{equation}

The most general state of a pure gauge theory is given by
\begin{align}
        \ket{\psi^{(2)}} = &\sum_{J_i, M_j, N_l} \alpha^{J_1J_2J_3j_aj_bj_cj_d}_{M_1M_2M_3N_1N_2N_3} \ket{J_1M_1N_1}\ket{J_2M_2N_2} \ket{J_3M_3N_3}.
\end{align}
We can substitute this state parametrization into Eq.~\eqref{eqn:Gauss_D8_trasformed}, where the solutions correspond to the states with eigenvalue $1$. The desired states are determined by the intersection of the solutions derived from the various Gauss' laws. As a result, we obtain the same 10 states identified in Section~\ref{solution of the trasformed model}, thereby concluding the proof. In conclusion, the evidence of the pseudo-2D analysis validates both the transformation and the Hamiltonian.

\subsection{Analytical verification of matter elimination: plaquette}\label{2d_plaquette_theory}

\input{figures/figure4}
In this case, constructing the gauge-invariant states for a plaquette as we did above in the case of the pseudo-2d model is significantly more involved. There are $158$ of them and finding the explicit form analytically is not feasible. 
Nevertheless, we can use the above certification of the $2$D model without plaquette term. 
The Hamiltonian of the plaquette can be extracted from the general $2$D formulation as a particular case with known boundary conditions. In this case, we also take into account the magnetic term, which we have verified to be correct as:
\begin{align}
\label{Magnetic_verification}
     &[W_g(\bold{x})\theta_{g,M}^\dag(\bold{x}),H_B] = 0 \\
     &[H_{\text{E}},H_{\text{B}}] = 0 = [H_{\text{M}},H_{\text{B}}] = 0 = [H_{\text{GM}},H_{\text{B}}]
\end{align}
By leveraging certain properties ($U$s and $\Pi$ are hermitian and real) derived from the gate formulation in Eq.~\eqref{U-definition}, the magnetic term can be rewritten as a sum of 16 terms, 
\begin{align}
    H_{\text{B}} = 2 Re[\lambda_b]\sum_{m, n, n', m'=0,1} \Pi^1 U^1_{mn}\Pi^2 U^2_{nn'}U^3_{m'n'}U^4_{mm'}.
\end{align}

The theory developed for a single plaquette, while simple, presents significant interest for implementation on real quantum hardware. 
Indeed, it describes the smallest non-trivial system that is already challenging to treat both analytically and computationally, thus offering a valuable testbed for benchmarking practical quantum-computing and -simulation algorithms.

\begin{figure*}
    \centering
    \includegraphics[scale = 0.35]{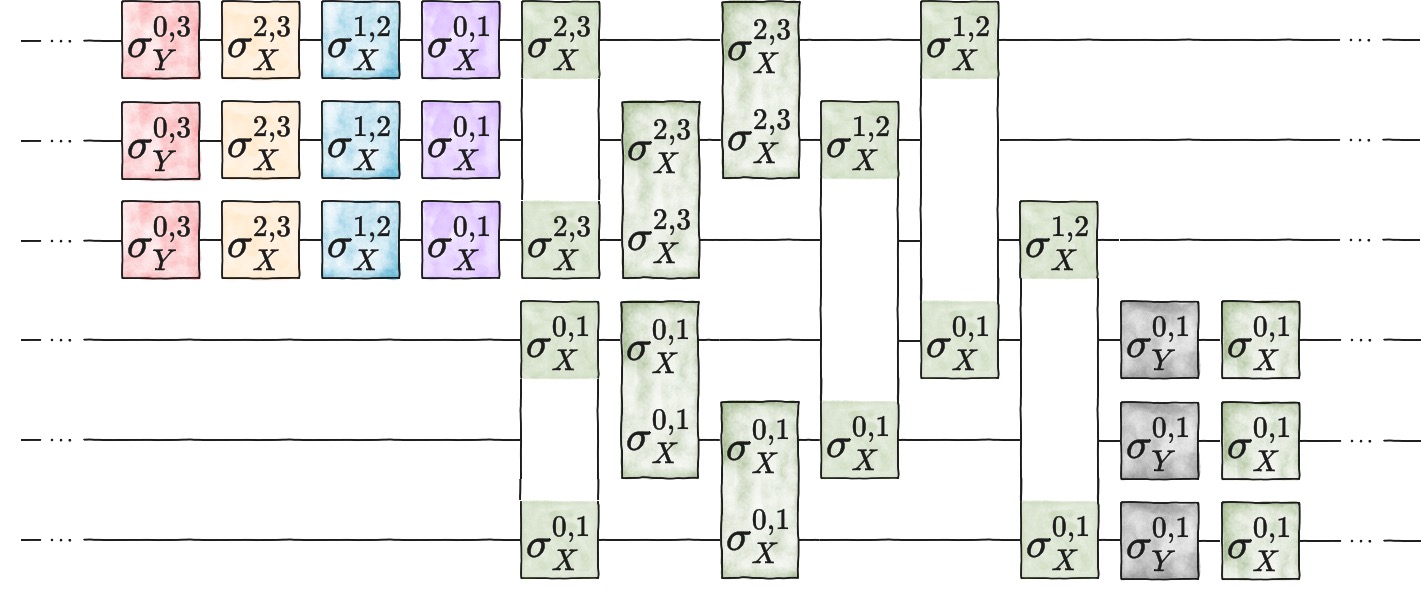}
    \caption{\textbf{Individual layer of the parametrized quantum circuit for the pseudo-2D system.} Each layer of the parametrized quantum circuit used to find the approximate ground state of the pseudo-2D system with non-Abelian $\mathbb{D}_8$-symmetry consists of two-level single-qudit rotations on the ququarts, entangling two-body MS gates between each pair of qubits and each pair of ququarts, non-local MS gates between a qubit and a ququart, and single-qubit Pauli-$X$ and Pauli-$Y$ rotations. Each gate in the variational circuit is parametrized by a variational parameter $\theta$, which differs between different layers.} 
    \label{fig:circuit_pseudo_2D}
\end{figure*}

\begin{figure*}[t!]
    \centering
\subfloat{\label{fig:pseudo_2D_energy}\includegraphics[width=0.5\textwidth]{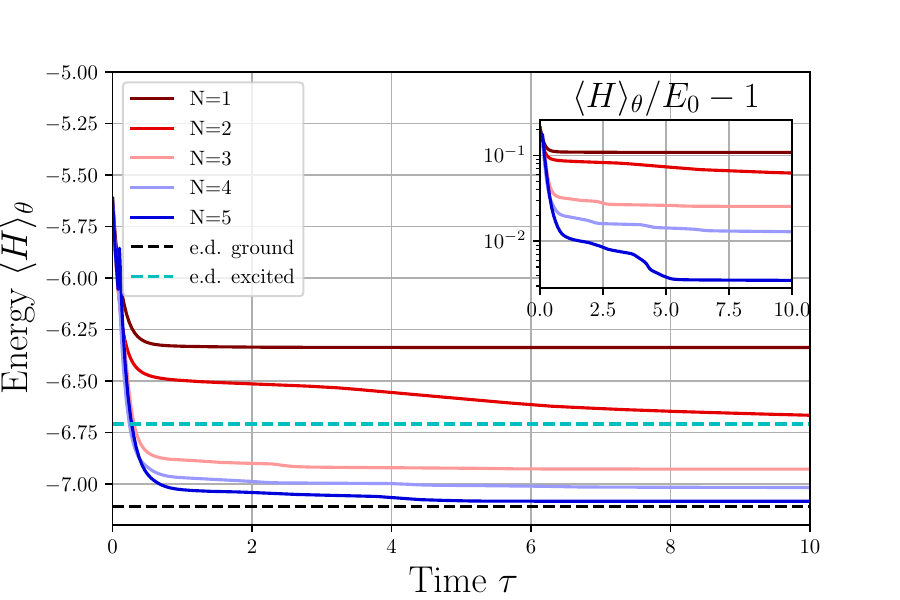}}%
\subfloat{\label{fig:pseudo_2D_fidelity}\includegraphics[width=0.5\textwidth]{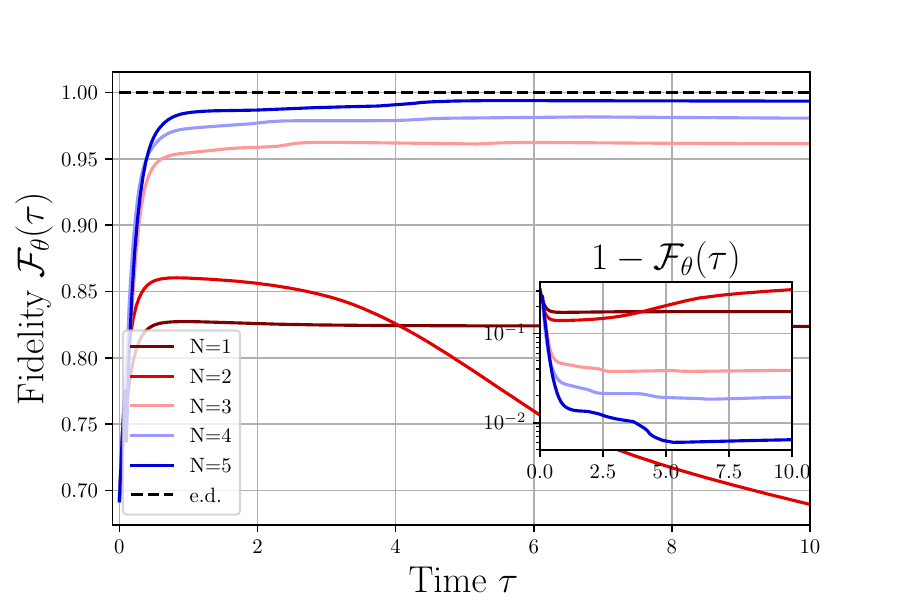}}%
    \caption{\textbf{VarQITE for the pseudo 2D system:} (a) The energy of the variational state $\ket{\psi(\boldsymbol{\theta)}}$ as a function of the imaginary time for $N = 1,2,3,4,5$ number of layers. The dashed black line shows the numerical value of the true ground state, whereas the cyan line - of the numerical first excited state. Note that the latter is not gauge invariant and depends on the coefficient $\lambda$ of penalty term introduced to the Hamiltonian, see Section \ref{sec:Simulations_pseudo_2D} (here, $\lambda = 0.05$). After initial fast convergence to low energies, the secondary slow optimization is primarily due to the penalty for Gauss' law violation. From the inset, one can deduce that after $\tau = 5$, the optimization practically stops, rendering relative error $<1\%$ for $N = 5$ layers. (b) The fidelity of the variational state w.r.t. the true ground state from exact diagonalization is shown as a function of the imaginary time $\tau$. As for the energy, the optimization practically stops after $\tau = 5$, rendering fidelity w.r.t. the ground state of $>99\%$ for $N = 5$ layers.}
    \label{fig:results_pseudo_2d}
\end{figure*}

\subsection{Simulations}
\label{sec:Simulations_pseudo_2D}

Here, we present the numerical results for the variational ground-state preparation of the pseudo 2D system. As in the case of the one-dimensional system, we again focus on the challenging point of the phase diagram $J = M = h = 1$, and employ the VarQITE algorithm in order to find an approximation of the ground state. Again, the variational circuit we use is composed of single-qubit, single-qudit, qubit-qudit and two-qudit gates that can be implemented in the trapped-ion qudit platform natively, ensuring the feasibility of our protocol. 
In FIG.~\ref{fig:circuit_pseudo_2D}, one layer of the circuit that implements the variational ansatz is shown. 

To initialize the VarQITE minimization, we choose the product state $\ket{\psi_0} = \bigotimes^3_{i = 1}\ket{1}_{b,i}\otimes\ket{3}_{d,i}$ and the variational parameters $\boldsymbol{\theta} = \boldsymbol{0}$. At every step of the imaginary time evolution, we calculate the energy of the variational state $E(\boldsymbol{\theta}) = \bra{\psi(\boldsymbol{\theta})}H\ket{\psi(\boldsymbol{\theta})}$ and the fidelity w.r.t.\ the exactly diagonalized ground state $\ket{\psi_{\text{ground}}}$ of $H$. Here, as in the case of the 1D system, we add to the Hamiltonian the penalty term $H \rightarrow H + \lambda\sum_i G_i^2$, where $G_i$ is the Gauss' law operator at site $i$ and $\lambda$ is a Lagrange multiplier.

In FIG.~\ref{fig:pseudo_2D_energy}, the course of minimization of the cost-function---the energy---is shown, and in FIG.~\ref{fig:pseudo_2D_fidelity} the fidelity with respect to the ground state. A fidelity over $95\%$ is reached already for $N = 3$ layers, whereas for $N = 5$ layers the fidelity is over $99\%$, corresponding to variational circuits with a total of $L_\mathrm{ent} = 3(4+4) = 24$ and $L_\mathrm{ent} = 5(4+4) = 40$ two-body entangling gates, respectively.

\label{2d_simulations}

%% file: figures/figure3.tex
\begin{figure}[h]
    \centering
    \begin{tikzpicture}
  \fill (-2,0) circle (3pt);
  \fill (0,0) circle (3pt);
  \fill (2,0) circle (3pt);
  \fill (0,2) circle (3pt);
\fill[blue] (-1,0) circle (3pt); 
\fill[blue] (0,1) circle (3pt); 
\fill[blue] (1,0) circle (3pt); 
 
  \node at (2,-0.4) {$x + e_1$}; 
  \node at (0, -0.4) {$x$}; 
  \node at (0,1.6) {$x + e_2$};
  \node at (-2,-0.4) {$x - e_1$};

    \node at (1.9,+0.4) {$c$}; 
  \node at (0, +0.4) {$d$}; 
  \node at (0,2.4) {$b$};
  \node at (-2,+0.4) {$a$};

  \node[blue]  at (-.4,1) {$2$}; 
  \node[blue]  at (-1, -0.4) {$1$}; 
  \node[blue] at (1,-0.4) {$3$};
\end{tikzpicture}
\caption{Pseudo-2D model, in blue the gauge field while in black the matter before being removed. We take $a, b$ and $c$ as the even sites while $d$ as the odd one. Per each site we have at most two fermions and they correspond to two different flavours; we are using the fundamental representation for the matter. The numbers will be used subsequently to reference the specific link under consideration.}
    \label{Figure3}
\end{figure}

%% file: figures/figure4.tex
\begin{figure}[h]
    \centering
    \begin{tikzpicture}

\fill[] (0,0) circle (3pt); 
\fill[] (3,3) circle (3pt);
\fill[] (0,3) circle (3pt);
\fill[] (3,0) circle (3pt); 
\fill[blue] (1.5,0) circle (3pt); 
\fill[blue] (0,1.5) circle (3pt); 
\fill[blue] (1.5,3) circle (3pt); 
\fill[blue] (3,1.5) circle (3pt); 
  \node at (2.9,-0.3) {$x + e_1$}; 
  \node at (0, -0.3) {$x$}; 
  \node at (0,2.7) {$x + e_2$};
  \node at (2.8,2.7) {$x + e_1 + e_2$};

  \node[blue]  at (-.4,1.5) {$4$}; 
  \node[blue]  at (1.5, -0.4) {$1$}; 
  \node[blue]  at (3.4,1.5) {$2$};
  \node[blue] at (1.5,3.4) {$3$};
\end{tikzpicture}
\caption{Single plaquette, in blue the gauge field and in black the matter before being removed. $x$ and $x + e_1 + e_2$ are taken as the even sites while $x + e_1$ and $x + e_2$ as the odd ones. Per site, we have at most two fermions corresponding to two different flavours. We defined the gauge Hilbert space as the product space of the links (1, 2, 3, and 4). }
    \label{Plaquette_figure}
\end{figure}

%% file: Sections/discussion_and_conclusions.tex
In this work, we have addressed two key challenges in the quantum simulation of lattice gauge theories (LGTs): the simulation of fermionic matter and the redundancy in the Hilbert space. By applying the transformation methods of Refs.~\cite{zohar_eliminating_2018, zohar_removing_2019} to the case of a $\mathbb{D}_8$ LGT with staggered fermions, we have demonstrated their effectiveness in simplifying non-Abelian theories (previously they were employed onlyin Abelian cases) and reducing resource requirements. As we have shown through analytical checks, the spectra of the system remain identical before and after the transformation, evidencing the faithfulness of the mapping. Further, by performing numerical simulations of the VarQITE algorithm employed for a parametrized quantum circuit on a trapped-ion qudit platform, we were able to show that using only an experimentally-friendly amount of resources in terms of gate count, the true ground state of 1D and 2D models with a $\mathbb{D}_8$ non-Abelian gauge symmetry can be approximated with high accuracy. Increasing the number of layers of alternating single-qudit and two-qudit entangling gates shows rapid improvements, both in terms of the value of the energy and in the fidelity with respect to the exactly diagonalized ground state. The relative error approaches $1\%$ for entangling gate count as low as 18 in the 1D case and 40 in the 2D case. 

Looking ahead, future research could focus on performing digital quantum simulations of the pseudo-2D model on actual quantum hardware or on extending the elimination techniques to more complex non-Abelian groups. This, in turn, could allow for the exploration of phenomena such as confinement and deconfinement, and potentially offer insights into the physical behavior of lattice gauge theories. From the implementation point of view, there is value in developing new tools that simplify quantum simulations, potentially by reducing degrees of freedom or minimizing required resources. Additionally, exploring methods for simulating compact Lie groups like $U(N)$ or $SU(N)$ may offer new directions for overcoming current experimental and technological limitations in quantum LGT simulations. Specifically, because these groups can give interesting insights into both strong and weak interactions.

%% file: Sections/appendixA.tex
In this section, we provide additional information regarding the matter removal process, going from the bosonic reformulation setting to the one without matter at all. This serves as an expanded explanation of the procedure introduced in Section~\ref{method2}. 

We start by studying how the single elements of the Hamiltonian terms are transformed under $\mathcal{V}^{(2)}$ defined in Eq.~\eqref{eqn:second_transf}), 
\begin{align} \label{eqn:U2_relations}
\mathcal{V}^{(2)}\tau^{\pm}_{n}\mathcal{V}^{(2)\dagger} &= P^{+}_n \tau^{\mp}_n + P^{-}_n \tau^{\pm}_n, \\
\mathcal{V}^{(2)}\tau^{z}_{n}\mathcal{V}^{(2)\dagger} &= -(P_m^+ - P_m^-) \tau_n^z,\\
\mathcal{V}^{(2)}U_{mn}\mathcal{V}^{(2)\dagger} &= \tau_m^xU_{mn}\tau_m^x.
\end{align}

\noindent Acting with $\mathcal{V}^{(2)}$ on the Hamiltonian, we find that the electric part is invariant,
\begin{equation}
	H^{(2)}_{\text{E}}=\mathcal{V}^{(2)}H^{(1)}_{\text{E}}\mathcal{V}^{(2)\dagger}= H^{(1)}_{\text{E}} = \lambda_E \sum_{l=links} \ket{2, mn} \bra{2, mn}.
\end{equation}
Recalling that $\Pi$ ($\equiv \theta_{a^2}$) is defined as the difference between the $j=2$ irrep projector and that for all the unfaithful irreps, we have
\begin{align}
    &\Pi = \Pi_1 - \Pi_2 \ \ \  , \ \ \ H_\text{E} = \lambda_E \sum_{l = links} \Pi_{2,l}.
\end{align} 
By omitting constant terms and rearranging, we ultimately obtain a term expressed solely in $\Pi$, whose corresponding gate form is known to be implementable \cite{ringbauer2022universal}, 
\begin{equation}
H^{(2)}_{\text{E}}  =-\dfrac{\lambda_E }{2}\sum_{l=links}  \Pi_l.
\end{equation}

Similarly to $H_E$, $H_M$ is also invariant. Nevertheless, we take a step back and consider its effective form. From the Gauss' laws in Eq.~\eqref{eqn:gauss_laws_with_number_operators}, we can write the mass term using solely gauge fields,  
\begin{align}
     H^{(2)}_\text{M} &= M \sum_{x,i} \varepsilon(\bold{x}) n_i(\bold{x}) = -\dfrac{M}{2} \sum_{x,i} W_i(\bold{x}).
\label{Mass_2d_term}
\end{align}
Recall that the $W_i(x)$, Eq.~\eqref{operator definitions}, consist of the operators $\theta_0$ and $\theta_1$, which again can be easily translated into gates Eq.~\eqref{theta-definitions}.
\input{figures/figure2}
The magnetic term $H_B$ remains invariant, as it is a pure gauge contribution. Its form is identical to that presented in Eq.~\eqref{hamiltonian terms after 1st trasf}. Therefore, $H_B$ involves only the gates corresponding to $\Pi$ and $U_{mn}$. 

Last, we examine the transformation of the interaction part, 
\begin{equation}
    H^{(2)}_{\text{GM}}=\mathcal{V}^{(2)}H^{(1)}_{\text{GM}}\mathcal{V}^{(2)\dagger}.
\end{equation}
Here, we need how terms like $\tau^+_m (\bold{x}) U_{mn}(\bold{x},1) \tau^-_n(\bold{x} + \hat{\mathbf{e}}_1)$ are transformed under $\mathcal{V}^{(2)}$, resulting in (using \fulleqref{eqn:U2_relations})
\begin{equation}
         \left(P^{+}_m \sigma^{-}_m + P^{-}_m \sigma^{+}_m\right)
 	\tau^+_m (\bold{x}) U_{mn}(\bold{x}) \tau^-_n(\bold{x} + e_1)
 \left(P^{+}_{n} \sigma^{+}_{n} + P^{-}_{n} \sigma^{-}_{n}\right).
\end{equation}
Reordering and using the properties of the $\tau$-Pauli matrixes, 
we observe that $H^{(2)}$ is block-diagonal in the matter spins, with static ($\tau_z = -1$) configurations. Using the constraints in Eq.~\eqref{eqn:gauss_law_method2}, we can restrict ourselves to physical states by integrating out all the matter-down states. Formally, \fulleqref{eqn:gauss_law_method2} implies that
\begin{equation}
	\left|\psi^{(2)}\right\rangle = \left|\tilde{\psi}^{(2)}\right\rangle \otimes \left|\Omega\right\rangle,
\end{equation}
where $\left|\Omega \right\rangle$ is a product state of all matter spins pointing down, and $\left|\tilde{\psi}^{(2)}\right\rangle$ is a state of the gauge fields. Then, we can define a Hamiltonian acting only on the gauge fields degrees of freedom via
\begin{equation}
	\tilde{H}^{(2)}= \left\langle \Omega \right| H^{(2)}\left|\Omega\right\rangle,
\end{equation}
including $\tilde{H}^{(2)}_{\text{M}} = H^{(2)}_{\text{M}}$, 
$\tilde{H}^{(2)}_{\text{E}} = H^{(2)}_{\text{E}}$, and $\tilde{H}^{(2)}_{\text{GM}}$.

The complete form of the final Hamiltonian becomes rather complicated. An explicit example is given in Sec.~\ref{sec:1d}. 
Assuming $J$ is real-valued (which can be done without loss of generality for $d=1$), the expression can be simplified by focusing exclusively on the anti-Hermitian components in the summation (like in \ref{summation}). As shown in FIG.~\ref{Figure2}, the radius of interactions is extended after the transformations $\mathcal{V}^{(2)}$ and $\mathcal{V}^{(1)}$, but, importantly, they nevertheless remain local. 



%% file: figures/figure2.tex
\begin{figure}[h]
\centering
   \begin{tikzpicture}
   \centering
  \scalebox{0.8}{\foreach \x in {2,4,6}
    \foreach \y in {0,2,4}
      \fill (\x,\y) circle (2pt); 
  \node at (1.9,1.8) {$x - e_1$};
  \node at (3.9,2.2) {$x$}; 
  \node at (3.9, -0.2) {$x - e_2$}; 
  \node at (5.9,2.2) {$x + e_1$};
  \node at (5.9,-0.2) {$x + e_1 - e_2$};
  \node at (3.9,3.8) {$x + e_2$};
  \node at (5.9,3.8) {$x + e_1 - e_2$};
  \node[blue] at (4,3) {$\theta_{a^2}$};
  \node[blue] at (3,2) {$\theta_{a^2}$};
  \node[blue] at (4,1) {$\theta_{a^2}$};
  \node[blue] at (4.9, 2.1) {$U_{00} \cdot \theta^R_{a^2y}$};
  \node[blue] at (7,2) {$\theta^L_{a^2y}$};
  \node[blue] at (6,3) {$\theta^L_{a^2y}$};
    \node[blue] at (6,1) {$\theta^R_{y}$};
    \node[red] at (4,1.7) {$\psi^\dag$};
    \node[red] at (6,1.7) {$\psi$};
    \node[red] at (5,1.7) {$U$};
    \fill[red] (4,2) circle (2pt);
    \fill[red] (6,2) circle (2pt);}
\end{tikzpicture}
\caption{Pictorial representation of an interaction term ($H^h_{00}(\bold{x})$). The vertices are empty as the matter fields have been eliminated. The blue operators act on the links (gauge d.o.f.); in red the original term.} 
\label{Figure2}
\end{figure}
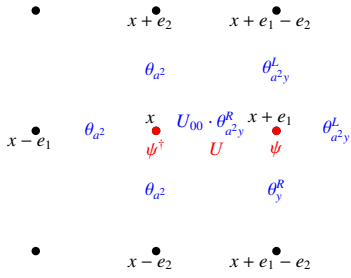